\documentclass[conference]{IEEEtran}
\ifCLASSINFOpdf
  \usepackage[pdftex]{graphicx}
  \graphicspath{ {./} }
\else
\fi
\usepackage{color}

\usepackage{url}

\begin{document}
%
\title{A Resource Management Protocol for Mobile Cloud Using Auto-Scaling}



%
\author{\IEEEauthorblockN{
Chathura Sarathchandra Magurawalage\IEEEauthorrefmark{1},
Kun Yang\IEEEauthorrefmark{1},
Ritosa Patrik \IEEEauthorrefmark{2},
Michael Georgiades\IEEEauthorrefmark{3} and
Kezhi Wang\IEEEauthorrefmark{1} 
}

\IEEEauthorblockA{\IEEEauthorrefmark{1}School of Computer Sciences and Electrical Engineering\\
University of Essex,
CO4 3SQ, Colchester, U.K.\\ Email: \{csarata, kunyang, kezhi.wang\}@essex.ac.uk}
\IEEEauthorblockA{\IEEEauthorrefmark{2}Telekom Slovenije, Ljubljana, Slovenia\\
Email: patrik.ritosa@telekom.si}
\IEEEauthorblockA{\IEEEauthorrefmark{3}PrimeTel PLC, Pafos, Cyprus\\
Email: michaelg@prime-tel.com}
}


\maketitle

\begin{abstract}
Cloud radio access networks (C-RAN) and Mobile Cloud Computing (MCC) have emerged as promising candidates for the next generation access network techniques. MCC enables resource limited mobile devices to offload computationally intensive tasks to the cloud, while C-RAN offers a technology that addresses the increasing mobile traffic. In this paper, we propose a protocol for task offloading and for managing resources in both C-RAN and mobile cloud together using a centralised controller. 



%
A working procedure of the proposed protocol has been illustrated to show it's mobile task offloading and resource allocation functions. The performance analysis of cloud vertical scaling indicates that the scaling time delay trend varies, depending on the scaling resource, i.e., CPU, RAM, Storage, in the mobile cloud. Furthermore, we conclude that using unaided vertical scaling for scaling computing resources in real-time, may not be suitable for delay sensitive applications, due to significant scaling delay.
\end{abstract}


%
\IEEEpeerreviewmaketitle

\section{Introduction}

User equipment (UE), e.g., smartphone, tablet, wearable device, and digital camera, is playing an important role in new application scenarios such as virtual reality, augmented reality and surveillance systems. While resource-constrained UE components such as CPU, GPU, memory, storage, and battery lifetime, have driven a dramatic surge in developing new paradigms to handle computation intensive tasks \cite{5445167}. For example, computationally intensive applications requiring a large amount of computing resources are not suitable to run on mobile or portable devices.


Mobile cloud computing (MCC) \cite{kumar2013survey} provides a solution where
the UE offloads computationally intensive tasks to a remote resourceful cloud (e.g. EC2), thereby saving processing power and energy. Furthermore, Kumar \emph{et al.} \cite{5445167} have discovered a computing-communication trade-off, and concluded that mobile task offloading is beneficial when the computing intensity of the task in question is high, and when the required network resources that are required to transfer the offloading task to the remote mobile cloud is relatively low. Kumar \emph{et al.} have also emphasised the need for high bandwidth wireless networks for task offloading to be efficient. 

\begin{figure}[htp]
	\includegraphics[scale=0.5]{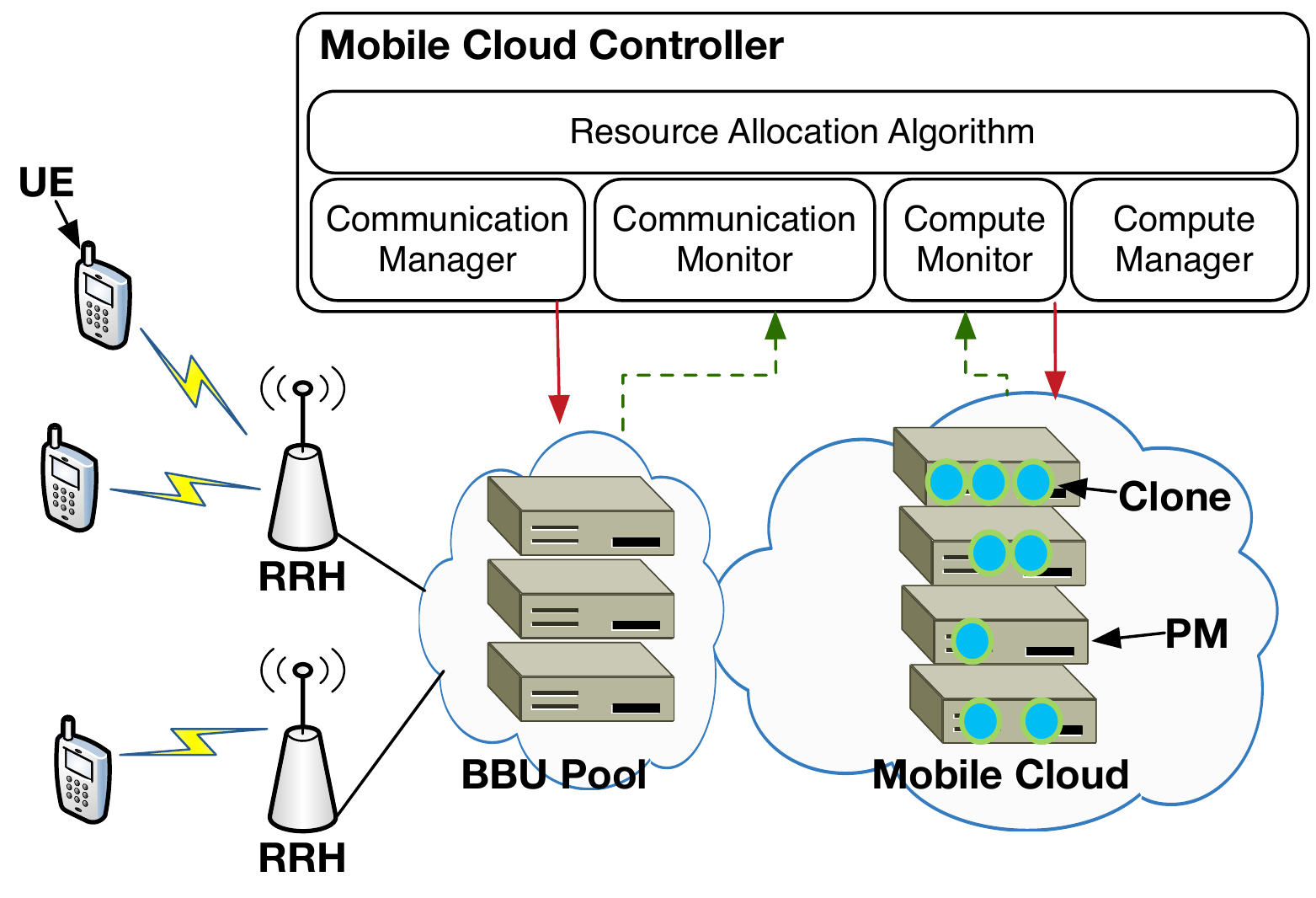}
	\caption{\label{chap:1:fig:C-RAN_MC}  Overview of the architecture, showing the interaction between, Mobile device and mobile cloud and C-RAN.}
\end{figure}

Subsequently, the offloading algorithm of the offloading framework that resides on the mobile device may make decisions on \emph{what} (which tasks), \emph{when} (when it is beneficial to offload) and \emph{where} to offload. Various authors have proposed different architectures for offloading frameworks and have provided their implementations. Examples of these include MAUI \cite{Cuervo:2010:MMS:1814433.1814441}, Thinkair \cite{6195845}, CloneCloud \cite{Chun:2011:CEE:1966445.1966473}, Cuckoo \cite{kemp2012cuckoo}. The architectures of which mostly dependent on the offloading type/level (thread level, method level, code level), and the implementation platform (programming language, used application libraries). However, the partitioning methods of task offloading are out of the scope of this paper. 


To ensure highly efficient network operation and flexible
service delivery when handling mobile internet traffic surging,
Cloud Radio Access Network (C-RAN) \cite{mobile2011c} brings cloud computing technologies into mobile networks by centralising baseband processing units (BBU) of the radio access network. It moves the BBU from traditional base station to the cloud and leaves the remote radio heads (RRH) distributed
geographically. The RRHs are connected to the BBU pool
via high bandwidth and low-latency fronthaul. The BBU pool
can be realised in data centres,
and the centralised baseband processing enables BBU to be
dynamically configured and shared on demand \cite{7105959}. In this
case, with the transition from conventional hardware-based
environment to a software-based infrastructure, C-RAN can
achieve flexible management of BBU resources.



Figure \ref{chap:1:fig:C-RAN_MC} shows the hybrid deployment
of C-RAN with a mobile cloud for computation offloading, as opposed to the traditional mobile cloud hosted on the internet. Connected
with geographically distributed RRHs and centralised BBUs,
UEs get access to the VMs (i.e., mobile clones) in
a mobile cloud for computation offloading. For computation
offloading requests, data is first transmitted by the base station
(RRH and BBU) Via the uplink. Once processed by a clone (Virtual Machine) in the mobile cloud, the results will be returned to UEs via the downlink. However, our work largely focuses on the mobile cloud side.

As shown in Figure \ref{chap:1:fig:C-RAN_MC}, a new kind of resource has been introduced to the traditional mobile operator's network. This is not only beneficial to the mobile users (UEs) for offloading computationally intensive tasks to the cloud, but there are a number of aspects that operators can benefit from. Operators are now not only a network pipe provider, but they can also offer computing services to the subscribers. Moreover, the operators may let the subscribers pay more on top of the current price plan for extra computing services that they receive, by introducing new price plans for mobile task offloading. One may introduce new components into existing systems, but such components still have to be managed for utilising the resources efficiently.

Software Defined Networking (SDN) \cite{6812298} is a centralised architecture that augments a data plane and a control plane from traditional networks. Centralised controller nodes have been introduced into the network for managing resources. Such SDN-based architectures and management controllers have been introduced to wireless networks in multiple accounts \cite{6845049} \cite{6702534}. The main focus of above has been on defining SDN modules and interfaces (northbound and southbound) for wireless networks \cite{6845049}, for efficient network service deployments. Moreover, SDN has been proposed for Long Term Evolution (LTE) wireless network control \cite{6845049}; e.g. interference mitigation, network access selection.

It is inescapable that a centralised controller needs to be introduced for managing both computing and resources in the mobile cloud and communication resources in C-RAN. Such a controller may still be able to control wireless network resources, while also managing computing resources when mobile tasks are offloaded to the cloud. Figure \ref{chap:1:fig:C-RAN_MC} depicts the proposed controller that manages the mobile cloud and baseband resources in C-RAN. The dotted lines illustrate the connectivity to both the BBU pool and the mobile cloud for sending control signals. In this architecture, a signalling protocol needs to be designed for task offloading and managing wireless resources in operator's network (Communication Manager) and computing resources in the mobile cloud (Compute Manager). However, this paper's focus is mainly on mobile task offloading, managing mobile cloud computing resources and on developing a prototype of such a system architecture to show further that the computational resources offloaded to the mobile cloud have to be managed efficiently while at the same time taking delay constraints of compute (task) offloading into account. 

The cloud resources in the mobile cloud have to be managed efficiently while also taking delay constraints of compute (task) offloading into consideration. Auto-scaling \cite{Zhan:2015:CCR:2775083.2788397} enables cloud administrators to dynamically scale computing resources in the cloud for their applications to adapt to workload fluctuations. There are two types of auto-scaling: 1) Horizontal scaling adds and removes virtual machines (VM) from an existing VM pool that serves an application, 2) Vertical scaling adds and removes virtual resources from an existing VM. However, horizontal scaling has been used more often, in comparison to vertical scaling in literature. Moreover, horizontal scaling allows the application to achieve higher throughput levels per each addition, but the deployment cost is greater than vertical scaling \cite{Dutta:2012:SAA:2353730.2353802}.

When scaling vertically, the resource provisioning introduces delay, which makes the desired effect arrive late. But not many previous auto-scaling techniques has taken these delays into consideration. \cite{Lorido-Botran:2014:RAT:2693546.2693559} stresses the need for future work on auto-scaling, taking auto-scaling time delays into consideration. Auto-scaling may scale either virtual disk, memory or CPU resources (virtual CPU is the most used auto-scaling resource) or a combination of types of resources. Also, one may scale up and scale down VMs depending on the current amount of workload. Moreover, when scaling resources, one may scale "continuously" by adding/removing just one resource from the existing VM or may add or remove more than one resource from the VM "non-continuously". The aforementioned scenarios should be taken into consideration when designing efficient offloading techniques. Therefore, it is important to understand the trends of such resize delays before designing auto-scaling algorithms, especially when auto-scaling for delay sensitive applications.

The remainder of this paper is organised as follows. Section \ref{sec:UOP} introduces the proposed task offloading and resource management, (i.e. unified) protocol. Then section \ref{sec:eval} presents the implemented prototype of the system described above and the performance analysis of the computing resource management operations using vertical auto-scaling. The conclusions are shown in section \ref{sec:conclusion} followed by the acknowledgement in section \ref{sec:ack}.

\section{Unified Protocol}
\label{sec:UOP}
Cloud Radio Access Networks allows cellular networks to process baseband tasks of multiple cells/RRHs as well as to allocate cellular resources to subscribers. In conventional mobile task offloading systems, the offloading framework and the mobile  network are not aware of the offloading process. The network treats the offloading data as any other data, and the offloading destination resides outside the network operator's network. The offloading process could have been more efficient if the offloading network resources and cloud resources can be dynamically allocated to fit offloading requirements. 


First, we design a protocol for communicating between the controller, BBU and the Mobile Cloud for resource allocation. This enables resource allocation when offloading. We assume that the UE has already discovered its BBU and its Clone, and propose a uniform PDU (Protocol Data Unit) format for both resource allocation and task offloading, as the name suggests. The proposed Uniform Offloading Protocol (UOP) operates in the Application Layer of the Internet Protocol Suite (OSI Layer 7). The Mobile Cloud Controller manages and is the main point of contact for both user (service consumer), and the services in the service providers side. We assume that the mobile cloud controllers are discoverable throughout the network for offloading using web service discovery protocols, e.g., Universal Description, Discovery and Integration. Once a mobile device discovers the most suitable mobile cloud to offload, it will then directly connect to the corresponding controller using the offloading protocol.

\subsubsection{Protocol Data Unit format}


\begin{figure}
	\centering
	\includegraphics[width=\linewidth]{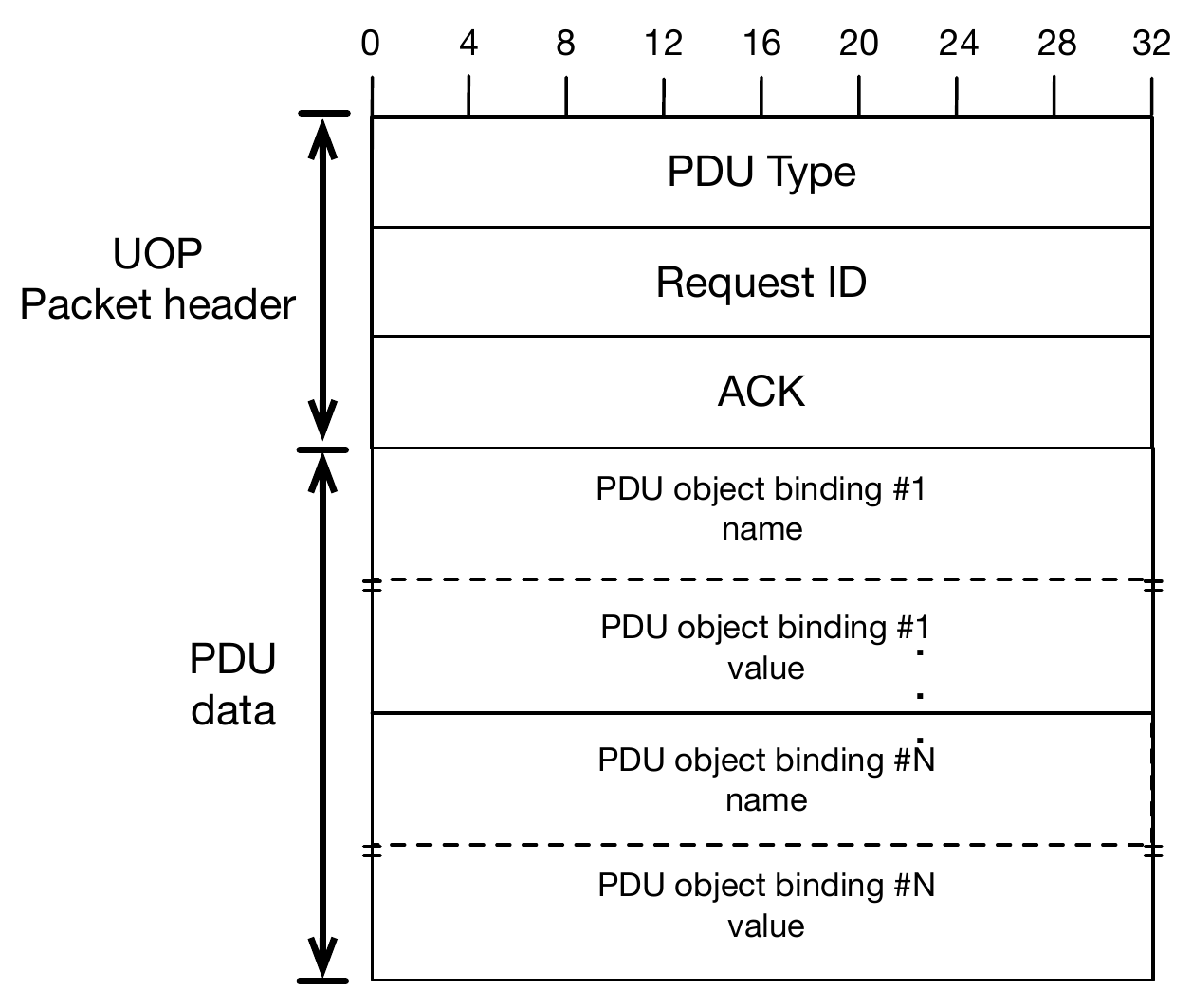}
	\caption{Unified Offloading Protocol PDU. (The ticks represent number of bits)}
	\label{fig:offloadpacketformat}
\end{figure}

\begin{table*}[]
	\centering
	\caption{UOP attribute definition}
	\label{tab:sec3:UOP_attribute_definition}
	\begin{tabular}{|l|l|l|l|}
		\hline
		Field Name                                               & Value Type       & Size (Bytes) & Description                                                                                                                                                                                                                                                                                                                       \\ \hline
		PDU type                                                 & Unsigned Integer & 4            & \begin{tabular}[c]{@{}l@{}}An integer value that indicates the type of PDU. \\ Refer to Table \ref{tab:sec3:pdutypes} for the list of PDU types.\end{tabular}                                                                                                                                                                                         \\ \hline
		Request ID                                               & Unsigned Integer & 4            & \begin{tabular}[c]{@{}l@{}}An identifier to match requests with replies. The \\ mobile device sets the Request ID in the request\\ PDU and then is copied by the controller and\\ the clone in the response PDU when offloading. \\ The controller sets this attribute when sending \\ resource management messages.\end{tabular} \\ \hline
		ACK                                                      & Unsigned Integer & 4            & \begin{tabular}[c]{@{}l@{}}0 if response packet, 1 if Acknowledgement packet \\ and, request rejection packet if the value \textgreater 1\end{tabular}                                                                                                                                                                                \\ \hline
		\begin{tabular}[c]{@{}l@{}}Object\\ Binding\end{tabular} & Object           & Variable     & \begin{tabular}[c]{@{}l@{}}A set of name-value pairs identifying application\\ objects to execute, user data and error messages\\ with their corresponding object references. Refer to\\ Table \ref{sec:3:definitions_of_object_binding_name-value_pair} for object definition.\end{tabular}                                                                                              \\ \hline
	\end{tabular}
\end{table*}

\begin{table}[]
	\centering
	\caption{PDU Types (some examples)}
	\label{tab:sec3:pdutypes}
	\begin{tabular}{|l|l|}
		\hline
		PDU Type value & PDU type        \\ \hline
		0000           & Offload\_Req    \\ \hline
		0001           & Offload\_Accept \\ \hline
		0002           & Offload\_Denied   \\ \hline
		0003           & Offload\_Start  \\ \hline
		0004           & App\_Register   \\ \hline
		0005           & App\_Request    \\ \hline
		0006           & App\_Data       \\ \hline
		0007           & App\_Response   \\ \hline
		0008           & Offload\_FIN    \\ \hline
		0009           & Manage\_Compute \\ \hline
		0010           & Manage\_BBU     \\ \hline
	\end{tabular}
\end{table}

\begin{table*}[]
	\centering
	\caption{Definitions of Object Binding name-value pair}
	\label{sec:3:definitions_of_object_binding_name-value_pair}
	\begin{tabular}{|l|l|l|l|}
		\hline
		Subfield Name & Value Type          & Size (Bytes) & Description                                                                                       \\ \hline
		Object\_Name  & Sequence of Integer & Variable     & \begin{tabular}[c]{@{}l@{}}Object identifier (code, user,\\ data, application error)\end{tabular} \\ \hline
		Object\_Value & Object              & Variable     & \begin{tabular}[c]{@{}l@{}}Contains the values of the \\ specified object type.\end{tabular}      \\ \hline
	\end{tabular}
\end{table*}

The PDU’s (Protocol Data Unit) format of both offloading and resource allocation protocol is shown in Figure \ref{fig:offloadpacketformat}. The offloading PDU is then passed down to the TCP layer and may be encapsulated with the TCP packet header. The offloading packet contains application control fields and a payload. The payload may contain offloading code, user data and application errors if any. Moreover, the definitions of the fields are shown in Table \ref{tab:sec3:UOP_attribute_definition}, the PDU types in Table \ref{tab:sec3:pdutypes} and the definitions of object binding name-value pair in Table \ref{sec:3:definitions_of_object_binding_name-value_pair}

We have avoided basing the offloading protocol on top of other web application protocols such as HTTP and SOAP due to their complexity and large overhead \cite{gray2005performance}. Henceforth, it operates on raw Transport Layer Sockets.  In our case, the offloading protocol is designed on top of TCP. Therefore, depending on the Transport Layer protocol that the offloading protocol (TCP/UDP) implements, the Client Handler in the controller and the offloading framework in the clone will listen on a known port for messages that are sent from the offloading mobile device. Implementing the protocol on top of TCP, automatically inherits TCP reliable delivery, congestion control, error correction and ability to add optional Transport Layer Security (TLS/SSL) layers. 

Time-outs takes an important role in the protocol for assuring the timeliness of individual transactions. Such timeouts are conventionally implemented within applications. Instead, the offloading protocol handles timeouts and indicates the applications if the protocol has timed out waiting for service responses. 

Each transaction (offloading or resource management procedure) must succeed, but in a case of a failure, the system should roll back to its previous state. The acknowledgement messages assure the completeness of transactions. If a task or a set of tasks fail, then an error message will be sent back instead of acknowledgement. The protocol assumes an offloading task as one transaction. Within this main transaction, there are multiple individual sub-transactions. Namely, they are communication resource allocation task, compute resource allocation task, remote code execution tasks. If any of aforementioned sub-transactions fail, the main transaction is also considered failed. The communication and computing resources that were allocated will be unallocated and put back into the pool of globally available resources. Finally, the code will have to be executed locally by the mobile device, if the delay restrictions do not allow it to resend an offloading request. From start to the end of an offloading transaction, mobile cloud controller keeps track of all protocol and application states.
 
We have kept the protocol and the packet format as simple as possible for reducing overhead and processing complexity. As depicted in Figure \ref{fig:initialisation_protocol} the protocol separates management functions (C-RAN/Mobile Cloud resource allocation) from services (task offloading) for increasing scalability and centralising management of services. It also Integrates application error reporting to the offloading protocol, (e.g. by including the error code in ACK section of the packet when the thinkair offloading framework has thrown an error) The offloading protocol is independent of the mobile operating system and the offloading framework. The payload of the packet can carry more than one offloading task.

\subsubsection{Working Procedure}

\begin{figure*}[]
	\centering
	\includegraphics[scale=0.3]{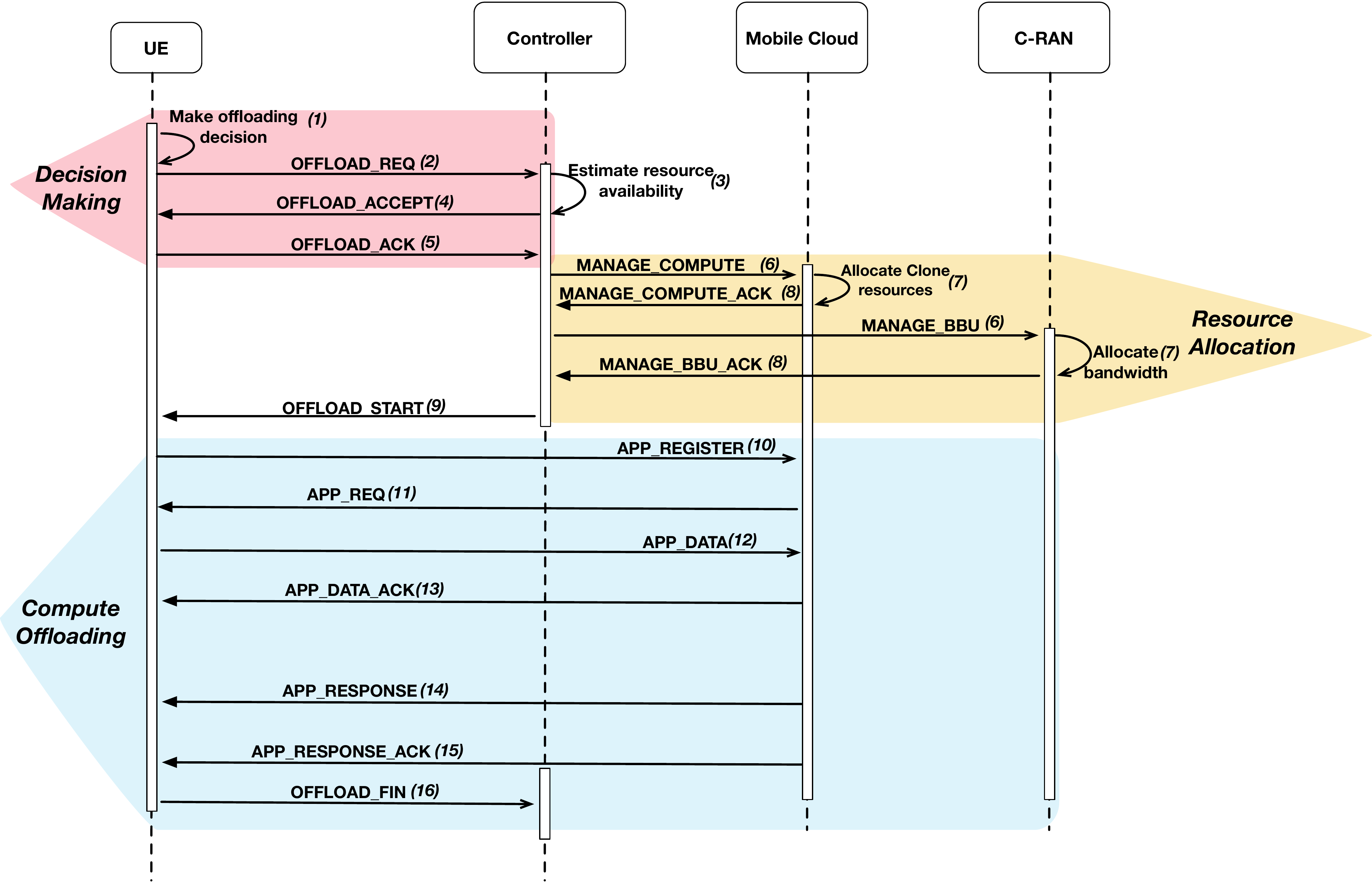}
	\caption{\label{fig:initialisation_protocol} Unified Offloading Protocol: Procedure when successfully allocates resources}
\end{figure*}


The Figure \ref{fig:initialisation_protocol} shows an instance where the protocol successfully instructs the UE to offload computationally intensive tasks to the mobile cloud. Once the UE receives the \emph{Offload\_Start} message with corresponding information about the offloading location (the clone), it successfully carries out offloading tasks. In the above figure, the acknowledgement messages are shown by appending ``\_ACK'' at the end of corresponding originating message name, to indicate the ACK bits has been set to 1 in the packet header. Resource monitoring is out of the scope of this protocol, and it is assumed that the controller uses existing monitoring protocols for monitoring C-RAN and mobile cloud resources. Moreover, C-RAN resource allocation, estimation and prediction algorithms are out of scope of this paper.

\section{Evaluation}
\label{sec:eval}
\subsection{Testbed Implementation}

\begin{figure}
	\centering
	\includegraphics[scale=0.7]{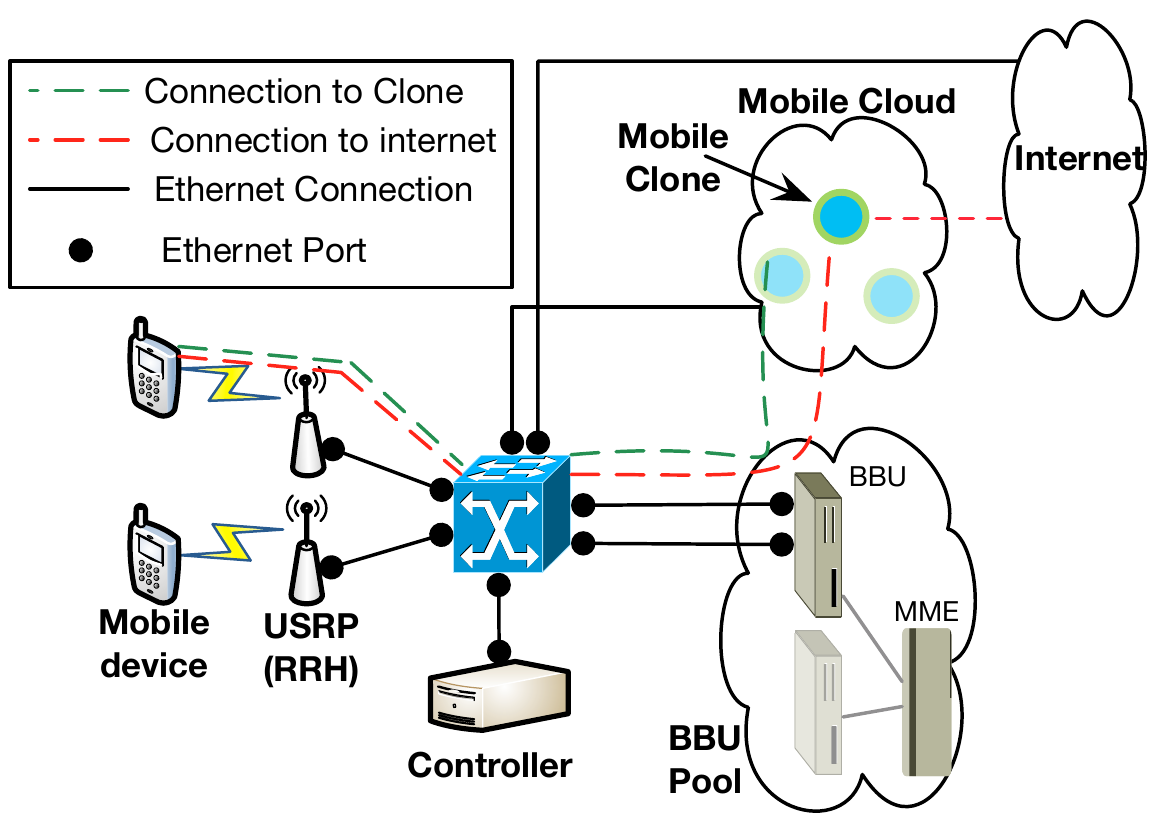}
	\caption{\label{fig:crantestbed} C-RAN with Mobile Cloud testbed}
\end{figure}

Figure \ref{fig:crantestbed}  shows the test environment set up for evaluating the proposed architecture. There are two USRP N210 and one X300 have been set up as RRHs of Amarisoft LTE 100 and OpenAirInterface (OAI) software base stations respectively. All nodes in the network are connected via a Gigabit Ethernet switch. The soft base stations are deployed on a Dell PowerEdge R210 rack server. OpenStack with Kernel Virtual Machine (KVM) as the hypervisor, has been deployed with no shared storage, for hosting the mobile clones running Android-x86 operating system. Thinkair \cite{6195845} has been used as the offloading framework, of which the server
components have been installed on the clone in mobile cloud,
while the client components are installed on the UE running
Android 4.4 operating system. The wireless bandwidth of the base stations has been set to 5 MHz. The BBUs are connected to its Mobility Management Entity (MME) via its S1-MME links using S1 Application Protocol (S1AP) that provides signalling between E-UTRAN and the evolved packet core (EPC). Host sFlow and sFlow-RT, monitoring and analytical tools, has been deployed on the controller as a part of the Resource Monitor for monitoring resources in the mobile cloud. We have also developed a monitoring module and a dashboard for monitoring wireless resources in the Mobile Cloud-RAN.

\subsection{Vertical Scaling for Resource Management in Mobile Cloud}


Prior to sending resource management instructions to the mobile cloud, using the proposed protocol, the resource management decisions are made in the controller. Auto-scaling in IaaS (Infrastructure as a Service) clouds has been studied extensively in the literature, for allocating computing resource optimally, while assuring Service Level Agreement (SLA) requirements and keeping the overall cost to a minimum. Although there exists an abundance of work on cloud auto-scaling, fundamental aspects of underlying technologies and algorithms that affect the performance of auto-scaling need improvement. Some known aspects of cloud computing that require improvements are \cite{Lorido-Botran:2014:RAT:2693546.2693559} \cite{Zhan:2015:CCR:2775083.2788397} \cite{7053814}, real-time hypervisors, real-time operating systems (OS), OS support for Cloudification (e.g. native cloud orchestration support in OSs), improved host resource sharing among guests (e.g. reduced latency), user friendly cost models and auto-scaling techniques for real-time applications. Specifically, when adopting cloud technologies into mobile systems, issues mentioned above are imperative due to high QoS/QoE requirements and dynamic nature of wireless networks and mobile applications \cite{6616117} \cite{Kumar2015191}. However, addressing aforementioned is out of scope of this paper. Our focus has been drawn to providing an extensive performance analysis of cloud vertical scaling, for improving auto-scaling algorithms, for real-time delay-sensitive applications (e.g. mobile task offloading) on existing cloud environments.

%
Cloud auto-scaling techniques are divided into predictive and reactive categories. Throughout literature authors have used various reactive and proactive techniques to do horizontal auto-scaling \cite{Bodik:2009:SML:1855533.1855545} \cite{CPE:CPE2864} \cite{6211900} \cite{6103960} \cite{5557965} by adding or removing VMs to/from a cloud application. Similarly vertical scaling \cite{6217477} \cite{Rao:2009:VRL:1555228.1555263} \cite{5071892} \cite{Shen:2011:CER:2038916.2038921} is done by adding or removing resources to/from existing VMs. Although throughout literature threshold rules based techniques seem to be most popular among authors, it can be seen that reinforcement learning, queuing theory, control theory, time series analysis have also been widely used. For both horizontal and vertical scaling, resource provisioning introduces a delay. Therefore the desired effect may arrive when it is too late. Current literature stresses the need for future work on auto-scaling focusing on reducing the time required to provision new VMs (or resize VMs when vertically scaling) \cite{Lorido-Botran:2014:RAT:2693546.2693559}. Moreover, comparatively horizontal scaling has been used predominantly in literature and by cloud service providers.

Vertical scaling is known to have a lower range when it comes to scaling \cite{Dutta:2012:SAA:2353730.2353802}, and for the changed resources to take effect, the VMs have to be restarted \cite{Lorido-Botran:2014:RAT:2693546.2693559}. Even if the underlying virtualization technology allows you to scale VMs without restarting (e.g. Xen CPU hot-plug), it may take up to 5-10 minutes for the changes to take effect and to be stabilised \cite{Rao:2009:VRL:1555228.1555263} \cite{6005367}. Moreover, this is because most conventional operating systems do not allow real-time dynamic configurations on the VMs without rebooting \cite{Lorido-Botran:2014:RAT:2693546.2693559}, and \cite{Rao:2009:VRL:1555228.1555263} reports this can also be partially due to the backlog of requests in prior intervals. However, in literature authors have proposed a number of vertical scaling algorithms, and they assume that vertical scaling actions can be performed timely, or the auto-scaling algorithms use hot-plugging \cite{5071892} \cite{6005367} \cite{Rao:2009:VRL:1555228.1555263} \cite{6217477}.  However, the latter is not possible with most of other popular hypervisors (e.g. Kernel Virtual Machine), cloud platforms (e.g. OpenStack) and cloud service providers (e.g. Amazon EC2 \cite{7463864}) \cite{Lorido-Botran:2014:RAT:2693546.2693559}. Due to above reasons, horizontal scaling has been more attractive to the community and has been adopted by existing cloud service providers instead of vertical scaling \cite{6217477} \cite{7463864}. Elastic Application Container (EAP) proposes an alternative provisioning mechanism to heavy VM provisioning \cite{6184989}, but it requires you to alter the underlying infrastructure. Furthermore some authors have proposed horizontal-vertical hybrid scaling systems where both approaches are used simultaneously for increasing performance, by utilising benefits of both methodologies \cite{6481061} \cite{6217477} \cite{6212070} \cite{5609349} \cite{Dutta:2012:SAA:2353730.2353802}.

%
%
%
%
It is considered wasteful to use a VM as the smallest resource unit when allocating computing resources to a  cloud application. The new VMs that are assigned to the applications are not used immediately after, and each VM consumes resources that are not directly utilised by the applications \cite{6217477}. As the number of VMs increase, the total number of resources that are consumed for just hosting and for keeping the VMs alive also increase. Furthermore, adding or removing a whole VM to an application is not always required for many real world scenarios \cite{6217477}, but subtle changes such as adding or removing available resources are sufficient. Dutta \emph{et al.} \cite{Dutta:2012:SAA:2353730.2353802} claims further that, while horizontal scaling allows the application to achieve higher throughput levels per each addition, the deployment cost is greater than vertical scaling. Therefore, this lead us to investigate further into performance trends of cloud vertical scaling.
The prototype described above has been left with out-of-the-box configurations for most software components (e.g. OpenStack), for making the evaluation results as generic as possible. The following analysis on VM creation and resizing times has been carried out by repeating each experiment 60 times for reducing noise due to other unavoidable influences within the environment. If the proposed protocol is used for vertical scaling, then the \emph{Manage\_Compute} PDU type can be used with the desired vCPU amount set in \emph{Object\_Binding} field of the packet (e.g. vcpu=4).

\begin{figure*}
	
	\centering
	
	\includegraphics[scale=0.3]{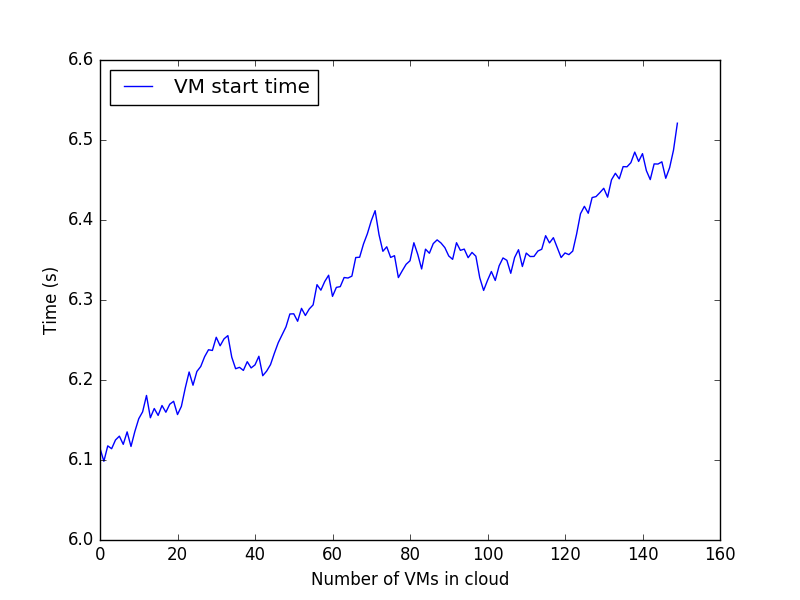}
	
	\caption{\label{fig:chap5:vmstarttimevstotalvms} The VM start time as the number of instances increases in the cloud globally}
	
\end{figure*}

There are two basic functions that one may perform on VMs when scaling either horizontally or vertically. One may instantiate and terminate new VMs behind a load balancer, or one would use a resizing function that is provided by the underlying cloud platform (hypervisor). Moreover, in this case, the Kernel Virtual Machine (KVM) environment allows the user to instantiate and terminate VMs. Figure \ref{fig:chap5:vmstarttimevstotalvms} depicts how the VM instantiation time has been influenced by all other VM created within the same cloud environment. We have applied a moving average function on the data for smoothing out the gathered results to highlight the increasing trend in data. The above graph clarifies the observations made by \cite{6253534}, where auto-scaling actions (horizontal) typically get delays in orders of minutes on public cloud service providers (Amazon EC2, Azure and Rackspace).  The above is mainly due to the increased request backlog and due to the time that it takes to assign a physical server to deploy a VM, then to move the VM image to it and get it fully operational. This further shows that instance start-time changes depending on the number of active VMs in the cloud at a given time. Moreover, when predicting the VM start times, knowledge on historical start delay times may help. We further stress that although the trend in data may be similar, the exact numbers at given points may change depending on the cloud environment.  Therefore, one may expect future cloud service providers to provide such data to be used by users when auto-scaling.

\begin{figure}
	\centering
	\includegraphics[scale=0.3]{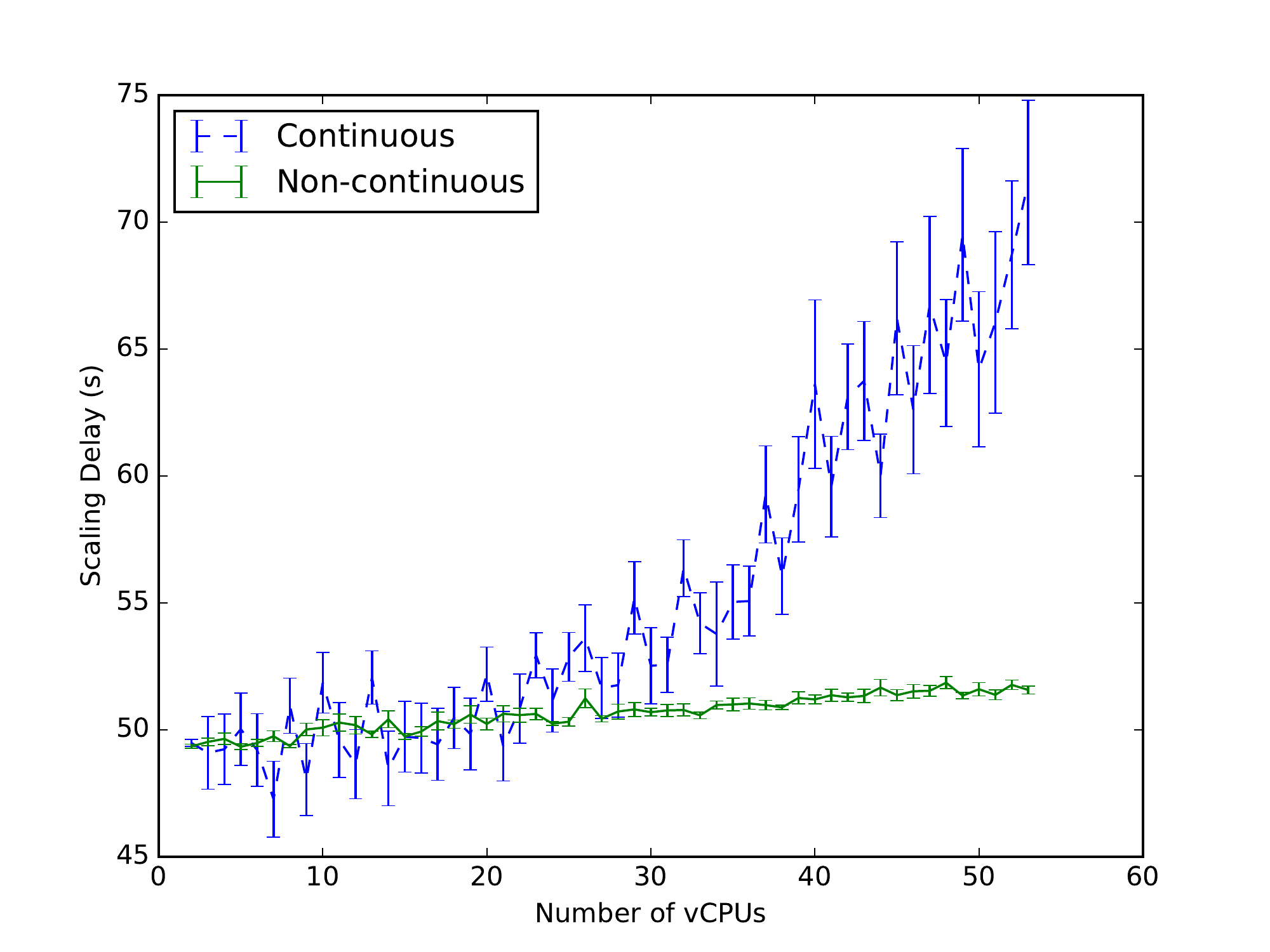}
	\caption{\label{fig:chap5:cpu_continuousVSstart} Mean CPU upscale delay as the size of the base VM increases. The standard error shows variations in results}
	
\end{figure}
\begin{figure}
	\centering
	\includegraphics[scale=0.3]{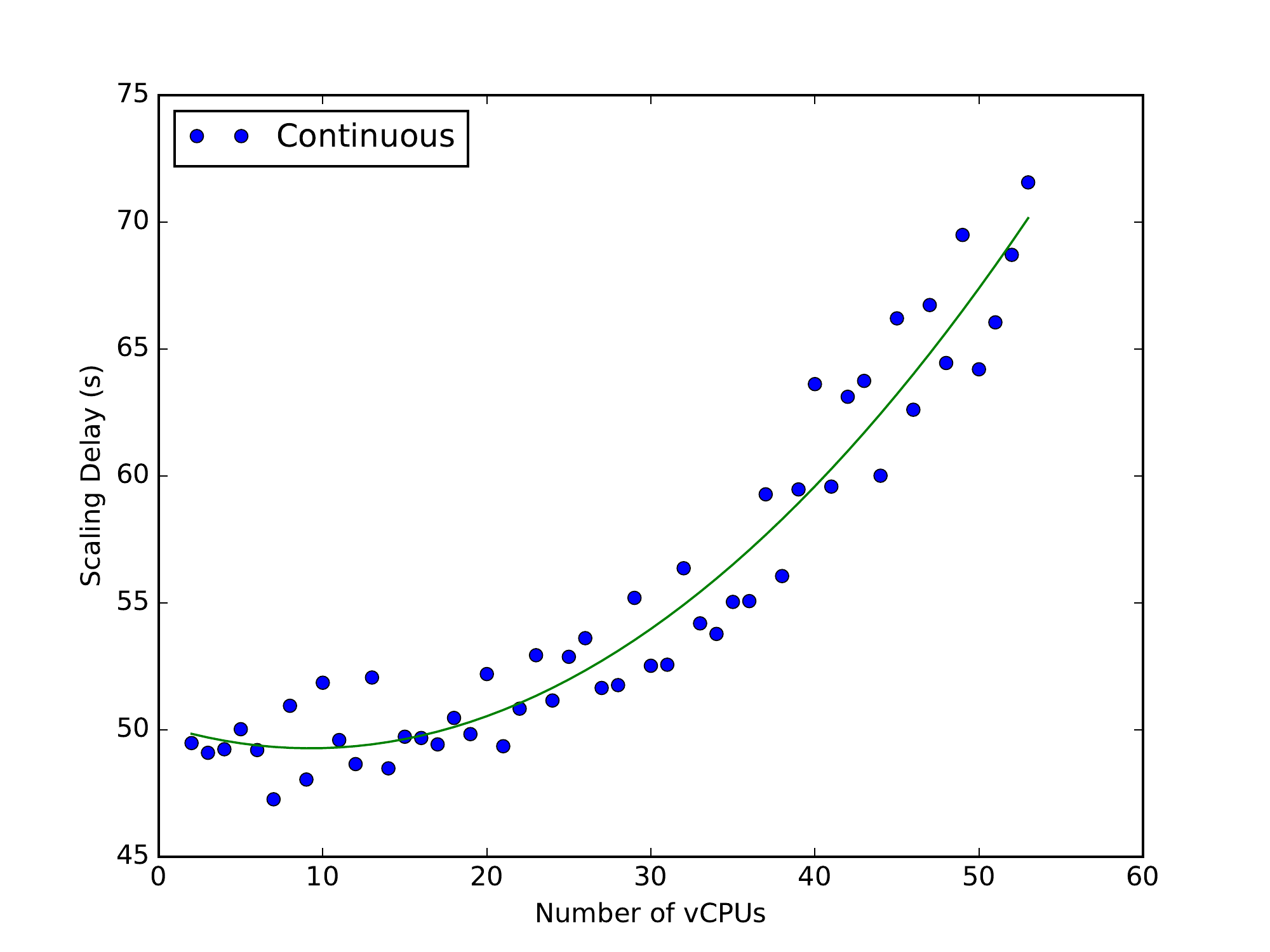}
	\caption{\label{fig:chap5:regressioncpu_continuous} Second order polynomial function of mean CPU upscale time in continuous scenario}
	
\end{figure}
\begin{figure}
	\centering
	\includegraphics[scale=0.3]{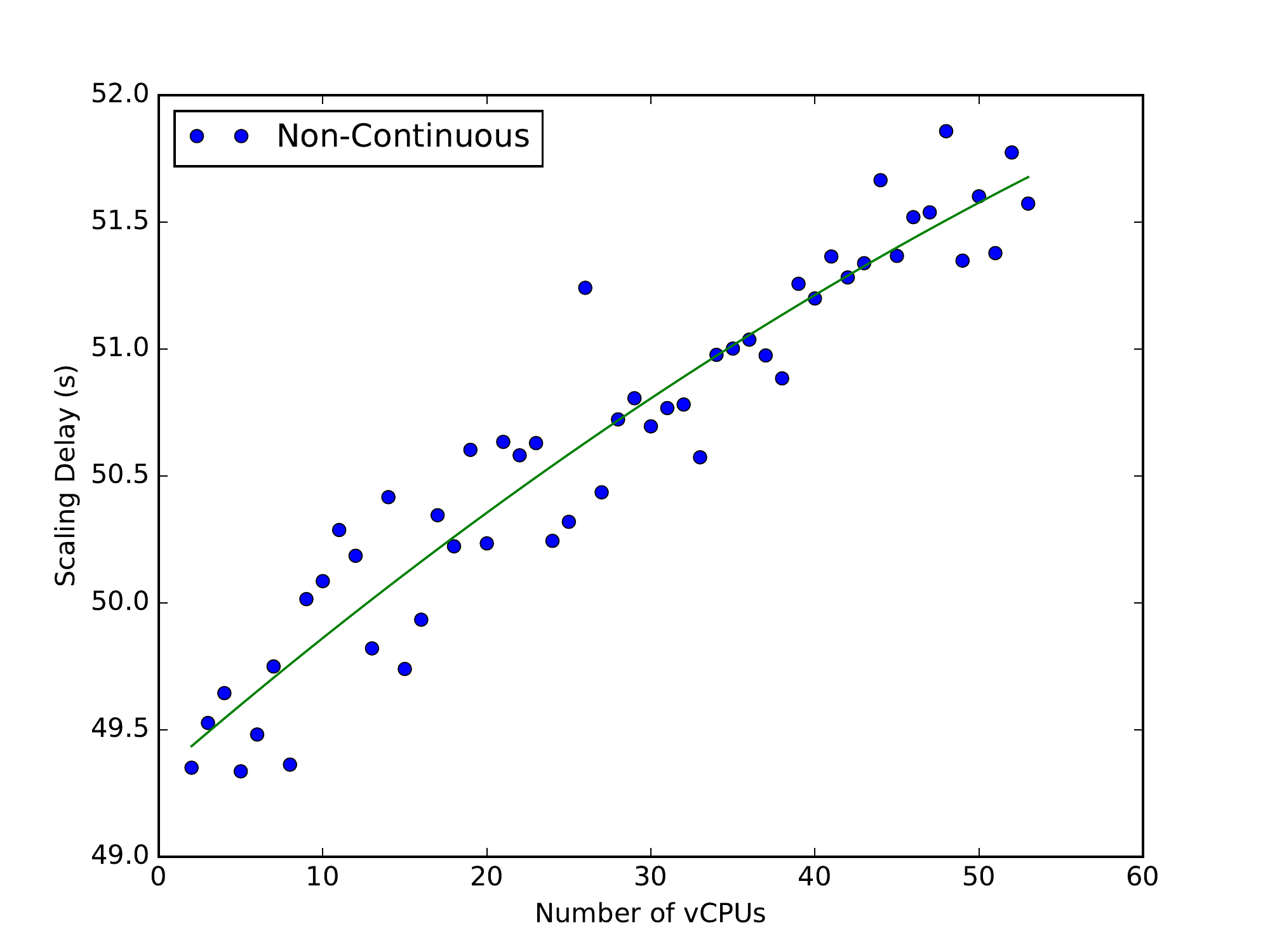}
	\caption{\label{fig:chap5:regressioncpu_startfrom1} Second order polynomial function of mean CPU upscale time in non-continuous scenario}
	
\end{figure}

We categorised auto-scaling into two scenarios. 1) Resizing continuously; the resources are added or removed by one at each iteration, 2) the resources are added or removed by amounts other than 1  (e.g. +1. +2, +3, +4). For the sake of the discussion, we call the latter "non-continuous". The former method may be suitable for algorithms that decide \emph{when} to scale (and later scales by adding/removing only one resource), and the latter method may be appropriate for algorithms that decide the \emph{amount} of resources to be added or removed at a given time. To avoid the after effects that may be caused by the previous iteration of auto-scaling commands, to the current iteration, we have added a 5-second sleep between each command.

We applied polynomial curve fitting \cite{bishop2006pattern} on the empirical data to further demonstrate how up and down scaling delay times vary across different types of resource and when different types of scaling are employed (continuously vs. non-continuously). Specifically, we have used a function of the form in equation \ref{sec:5:least_square_ploly} for fitting the data, where $M$ denotes the order of the polynomial function, given a training data set comprising $N$ observations of $x$, where $x \equiv (x_1,\dots,x_N)$ and corresponding observations $t \equiv (t_1,\dots,t_N)$. The vector $W$ contains the polynomial coefficients $w_0,\dots,w_M$.

\begin{equation}
\label{sec:5:least_square_ploly}
y(x, W) = w_0 + w_1x + w_2x^2 + \dots + w_Mx^M = \sum_{j=0}^{M} w_jx^j
\end{equation}

The coefficients are calculated by fitting the polynomial to the provided scaling delay data by minimising the squared error $E(W)$, as shown in equation \ref{sec:5:least_square_ploly_error}. 

\begin{equation}
\label{sec:5:least_square_ploly_error}
E(W) = \sum_{n=0}^{N} | y(x_n, W) - t_n|^2
\end{equation}

where $t_n$ denotes the corresponding target values for $x_n$. It measures the misfit between the function $y(x, W)$, for any given value of $W$ and scaling delay data points. We have calculated the second order polynomial function using least square polynomial fit \cite{bishop2006pattern} for the data, only for illustrating trends in data and to show differences in values, as shown in second order polynomial equation \ref{sec:5:secondorder_reg}. Such a model can be adapted to the architecture proposed in this paper where the polynomial function is used for predicting resize delays in the mobile cloud controller (when making auto-scaling decisions), the order of the polynomial function should be chosen to fit the data best. For providing further insight into the analysis, we have provided the polynomial coefficients in Table \ref{sec:5:polynomial_coefficients} that can be used with the equation \ref{sec:5:secondorder_reg} for evaluating vertical scaling algorithms in future. In equation \ref{sec:5:secondorder_reg} the $w_2$, $w_1$, $w_0$ coefficients can be looked up from the Table \ref{sec:5:polynomial_coefficients}, while $x$ denotes the amount of resources to evaluate on.

\begin{equation}
\label{sec:5:secondorder_reg}
y(x) = w_2x^2 + w_1x + w_0
\end{equation}


\begin{table}[]
	\centering
	\caption{Polynomial coefficients of second order polynomial function for the scaling scenarios}
	\label{sec:5:polynomial_coefficients}
	\begin{tabular}{l|l|l|l|}
		\cline{2-4}
		& \multicolumn{3}{c|}{\textbf{Coefficients}} \\ \hline
		\multicolumn{1}{|l|}{\textbf{Scaling Scenario}}                                    & $w_2$        & $w_1$     & $w_0$  \\ \hline
		\multicolumn{1}{|l|}{Figure \ref{fig:chap5:regressioncpu_continuous}}     & 0.0109       & 0.2013    & 50.2   \\ \hline
		\multicolumn{1}{|l|}{Figure \ref{fig:chap5:regressioncpu_startfrom1}}     & -0.0002161   & 0.05584   & 49.32  \\ \hline
		\multicolumn{1}{|l|}{Figure \ref{fig:chap5:rev_regressioncpu_continuous}} & 0.01358      & 0.3637    & 51.61  \\ \hline
		\multicolumn{1}{|l|}{Figure \ref{fig:chap5:rev_regressioncpu_startfrom1}} & 0.0003889    & 0.04552   & 66.85  \\ \hline
		\multicolumn{1}{|l|}{Figure \ref{fig:chap5:regressiondisk_continuous}}    & -1.159e-05   & 0.1038    & 46.92  \\ \hline
		\multicolumn{1}{|l|}{Figure \ref{fig:chap5:regressiondisk_startfrom1}}    & 2.837e-05    & 0.008834  & 47.05  \\ \hline
		\multicolumn{1}{|l|}{Figure \ref{fig:chap5:regressionmem_continuous}}     & -0.002184    & 0.04266   & 49.07  \\ \hline
		\multicolumn{1}{|l|}{Figure \ref{fig:chap5:regressionmem_startfrom1}}     & 0.007889     & 0.1701    & 50.31  \\ \hline
		\multicolumn{1}{|l|}{Figure \ref{fig:chap5:rev_regressionmem_continuous}} & 0.1402       & 2.375     & 56.91  \\ \hline
		\multicolumn{1}{|l|}{Figure \ref{fig:chap5:rev_regressionmem_startfrom1}} & 0.03394      & 0.6107    & 53.26  \\ \hline
	\end{tabular}
\end{table}

CPU resizing has been more often used than resizing other resources. Figure \ref{fig:chap5:cpu_continuousVSstart} shows that the time varies when adding resources, depending on the vCPU amount it increases from (i.e. the number of vCPUs that the VM has before resizing), in continuous auto-scaling scenario. Another observation that we can gather from the graph is that the VM resizing time increases as the size of the VM increases. One may observe that the change in results in the non-continuous scenario is much smaller as the scale of the VM increases, albeit there is still an increasing trend. The aforementioned statements are further clarified in Figure \ref{fig:chap5:regressioncpu_continuous} and Figure \ref{fig:chap5:regressioncpu_startfrom1} respectively. Both Figures \ref{fig:chap5:regressioncpu_continuous} and \ref{fig:chap5:regressioncpu_startfrom1} show second order regression functions of results for continuous and non-continuous scenarios respectively, depicting that time delay increases as the VM size increases. The graph also shows the standard error of each point to show how results varied in the results. 

\begin{figure}
	\centering	
	\includegraphics[scale=0.3]{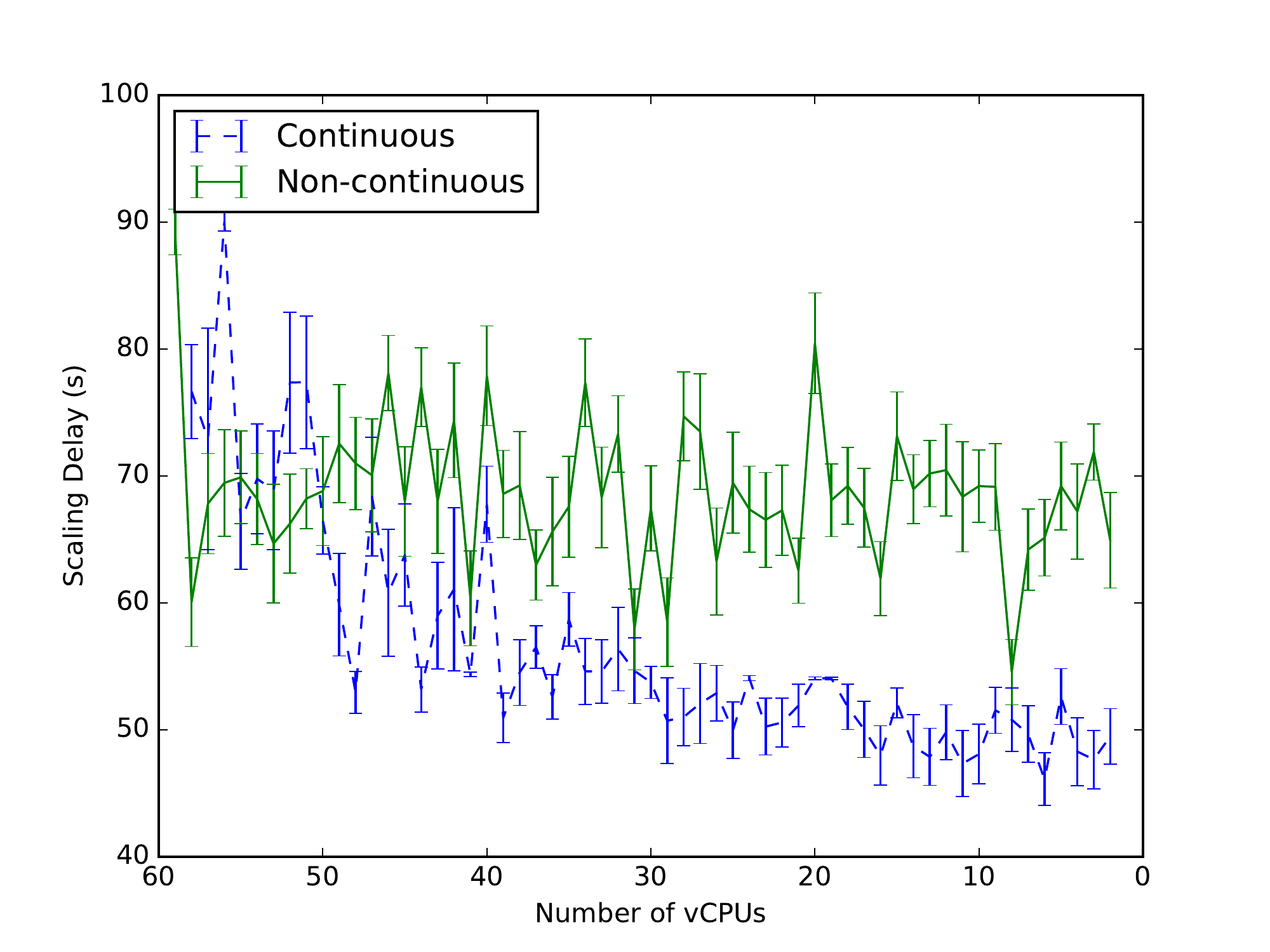}
	\caption{\label{fig:chap5:rev_cpu_continuousVSstart} Mean CPU downscale delay as the size of the base VM increases. The standard error shows variations in results}		
\end{figure}

\begin{figure}
	\centering
	\includegraphics[scale=0.3]{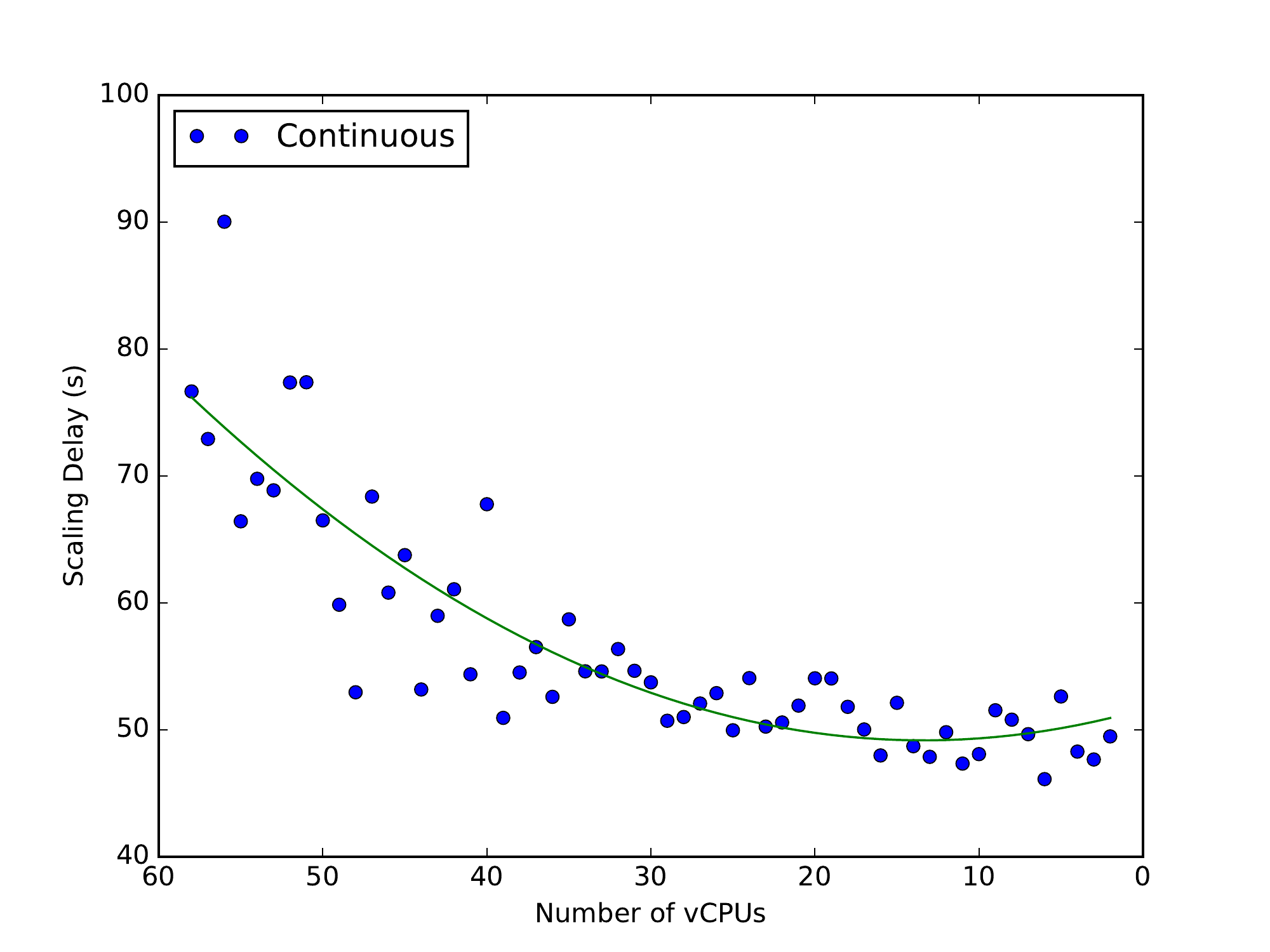}	
	\caption{\label{fig:chap5:rev_regressioncpu_continuous} Second order polynomial function of mean CPU downscale time in continuous scenario}		
\end{figure}

\begin{figure}
	\centering
	\includegraphics[scale=0.3]{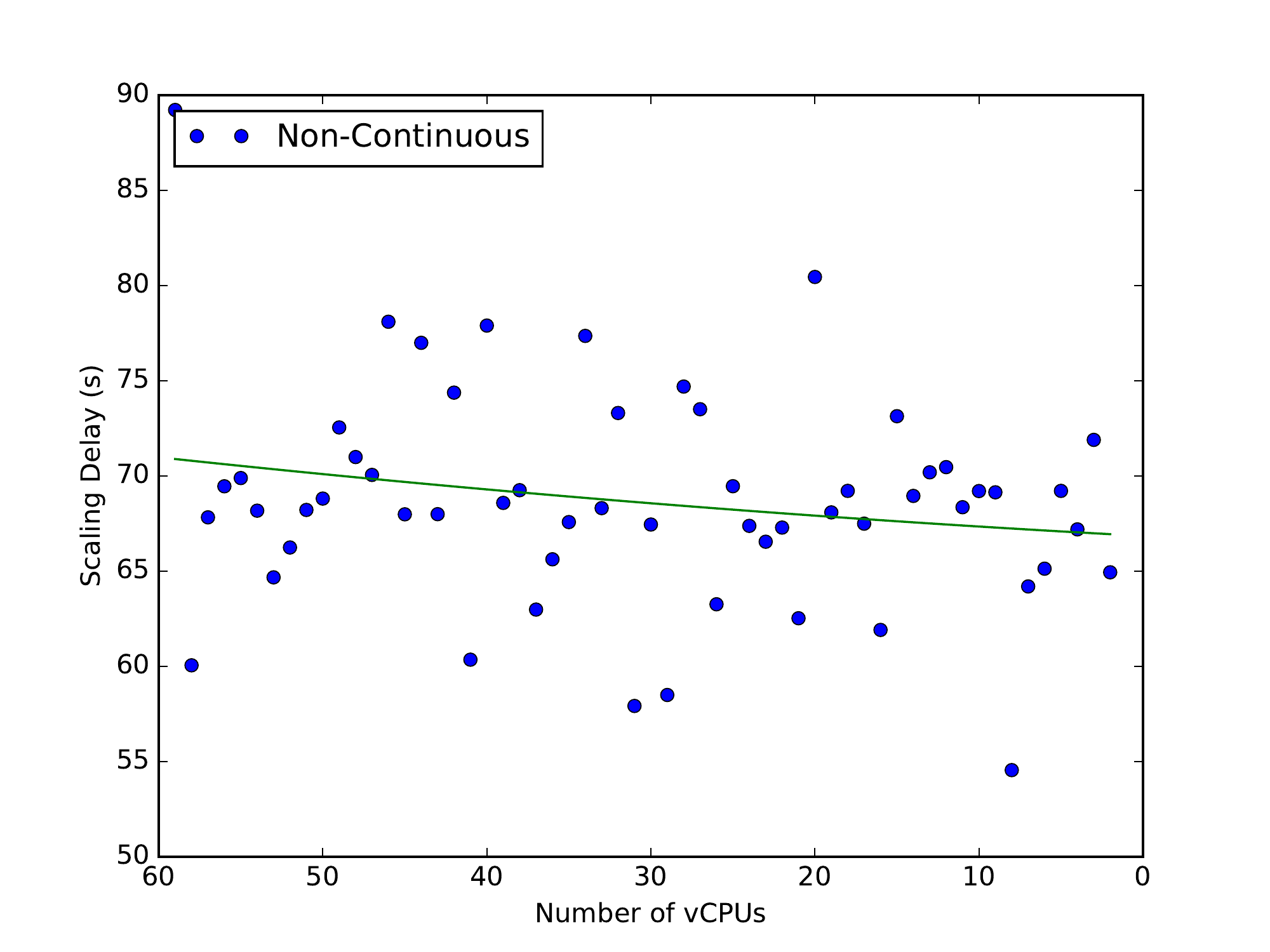}
	\caption{\label{fig:chap5:rev_regressioncpu_startfrom1} Second order polynomial function of mean CPU downscale  time in non-continuous scenario}
\end{figure}

Similarly to VM CPU upscaling, Figure \ref{fig:chap5:rev_cpu_continuousVSstart}, Figure \ref{fig:chap5:rev_regressioncpu_continuous} and Figure \ref{fig:chap5:rev_regressioncpu_startfrom1} shows continuous and non-continuous cloud performances of CPU downscaling. Specifically, the Figure \ref{fig:chap5:rev_cpu_continuousVSstart} depicts a comparison of delay times when virtual CPUs are scaled down continuously and non-continuously. The error bars denote the standard error at each point to show the variations of gathered results. One may observe that this has an opposite trend (decreasing trend) to VM CPU upscaling delay as shown in Figure \ref{fig:chap5:cpu_continuousVSstart}, where the delay time decreases as more CPUs are removed from the VM. One may conclude from the analysis that the resize time is greatly influenced by the size of the VM when scaling computing resources (i.e. the number of CPUs the VM has before resizing). Moreover, the Figure \ref{fig:chap5:rev_regressioncpu_continuous} and the Figure \ref{fig:chap5:rev_regressioncpu_startfrom1} shows second order polynomial function of both continuous and non-continuous mean CPU downscaling delay in cloud respectively. Furthermore, the observations suggest that when designing auto-scaling algorithms that scale CPU resources, the future work should consider the time differences when scaling continuously and scaling non-continuously.

\begin{figure}
	\centering
	\includegraphics[scale=0.3]{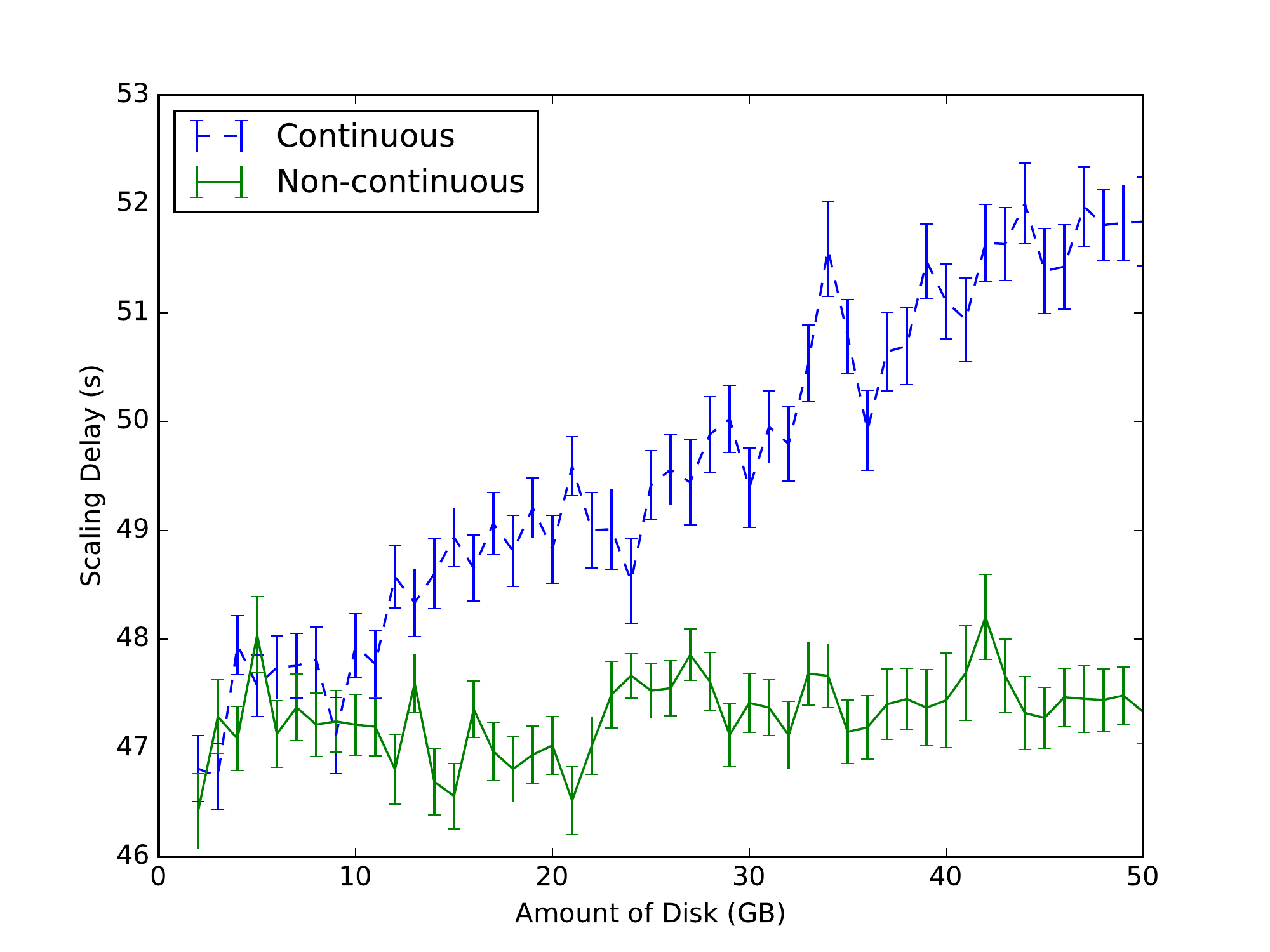}
	\caption{\label{fig:chap5:disk_continuousVSstart} Mean disk upscale delay as the size of the base VM increases. The standard error shows variations (error) in data}
	
\end{figure}
\begin{figure}
	\centering
	\includegraphics[scale=0.3]{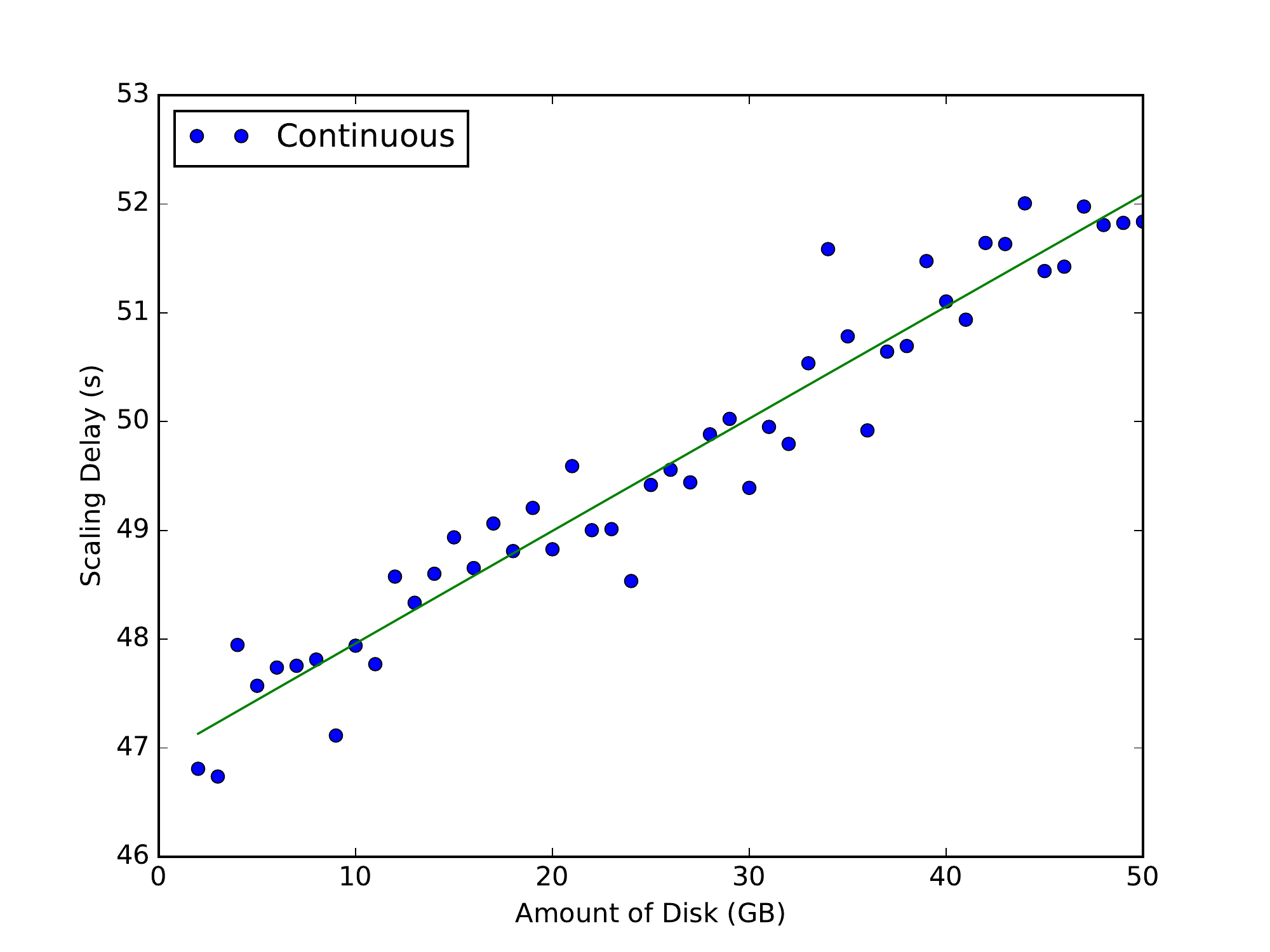}
	\caption{\label{fig:chap5:regressiondisk_continuous} Second order polynomial function of mean disk upscale time in continuous scenario}
	
\end{figure}
\begin{figure}
	\centering
	\includegraphics[scale=0.3]{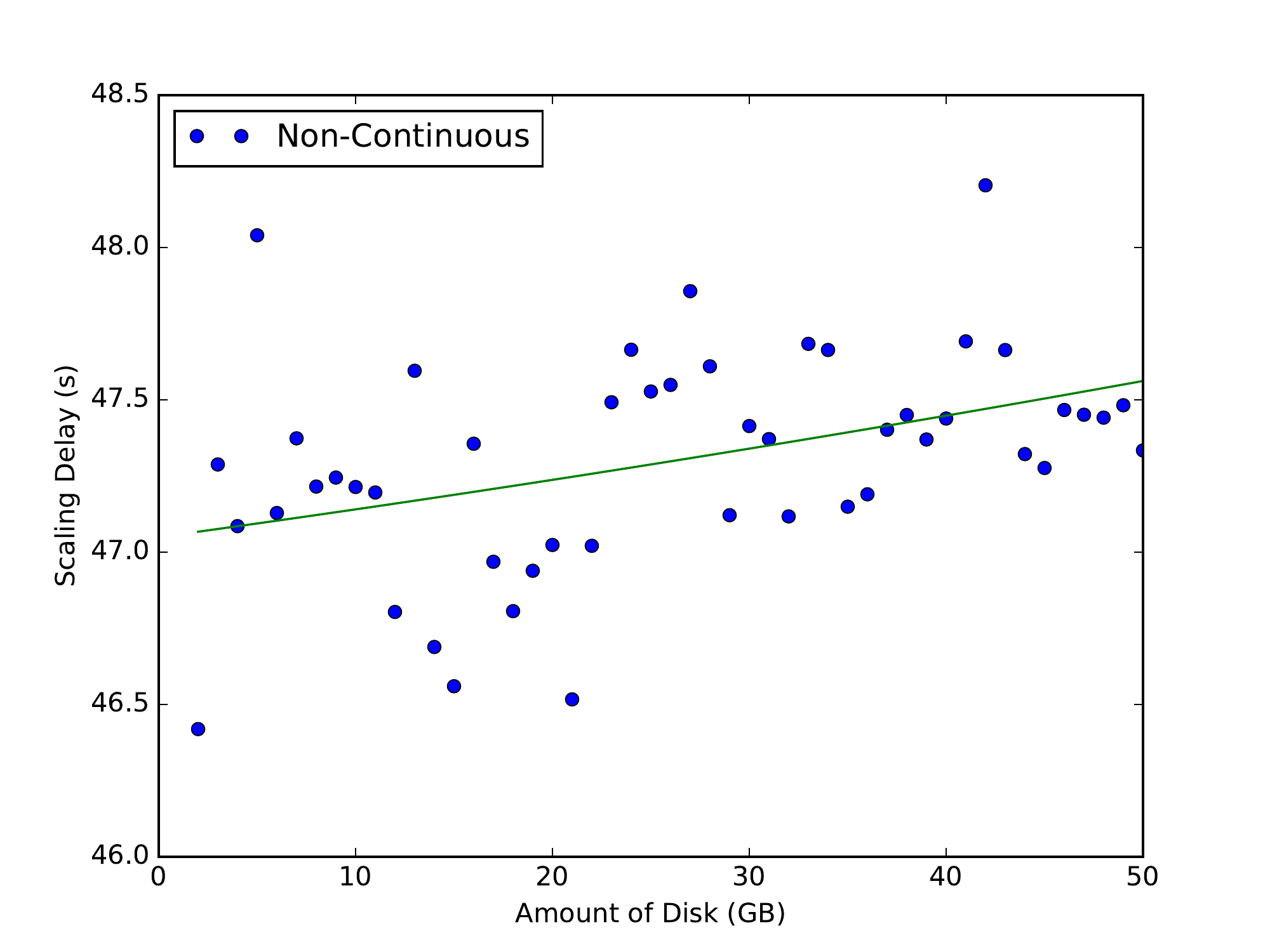}
	\caption{\label{fig:chap5:regressiondisk_startfrom1} Second order polynomial function of mean disk upscale time in non-continuous scenario}
	
\end{figure}

According to our experiments, the second most influential resource on the resize time delay is the VM disk size. However, in the used test environment, disk downscaling is not supported by the cloud framework (OpenStack with KVM). Therefore only the upscaling results are presented. The Figure \ref{fig:chap5:disk_continuousVSstart} shows an opposite trend to the CPU resize time delays in Figure  \ref{fig:chap5:cpu_continuousVSstart}. In this experiment, the non-continuous resizing time delays appear to be higher than when resizing continuously. Moreover, the resize time does not show any significant increases or decreases as the amount of disk space increase in the non-continuous scenario. A second order polynomial regression analysis has been carried out in Figure \ref{fig:chap5:regressiondisk_startfrom1} further clarifying the above observation. 

It is clear that the resize delay time increases as the VM's disk size increases in the continuous VM scaling scenario as shown in Figure \ref{fig:chap5:regressiondisk_continuous}. Such an observation is expected, as when resizing, some hypervisors (e.g. KVM) take a snapshot of the running VM, then a new resized VM is created from the snapshot with changed configurations (often on a different compute node). Then the old VM gets deleted. Whereas, when we have conducted the non-continuous auto-scaling experiment, we chose 1GB as the base VM disk size. Therefore, one can expect the effect of the base disk size when resizing at each step to be even. Moreover, the base disk size stays constant when scaled non-continuously, although the VM is resized to a different disk size.

\begin{figure}
	\centering
	\includegraphics[scale=0.3]{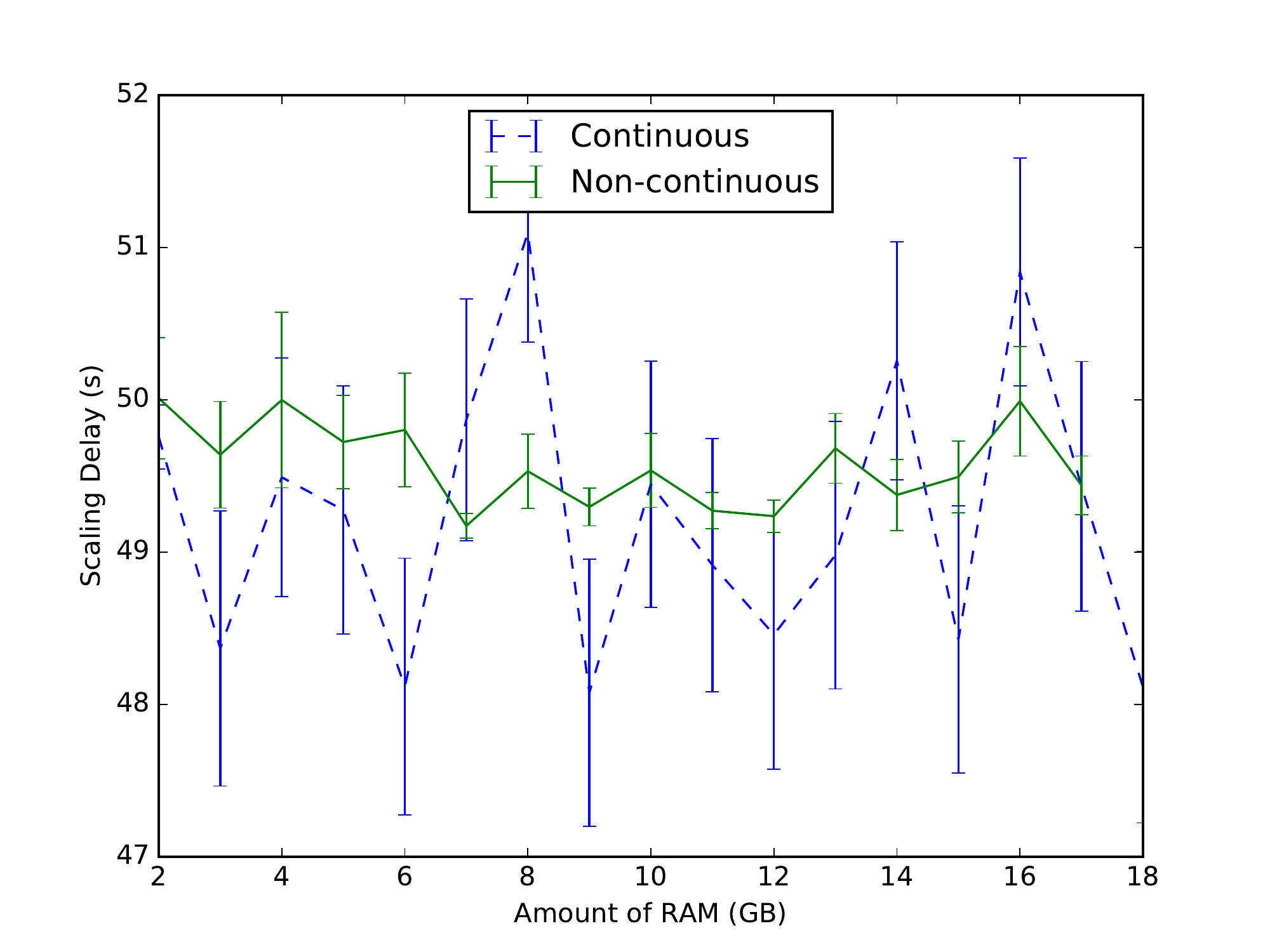}
	\caption{\label{fig:chap5:mem_continuousVSstart} Mean RAM upscale delay as the size of the base VM increases. The standard error shows variations (error) in data}
	
\end{figure}
\begin{figure}
	\centering
	\includegraphics[scale=0.3]{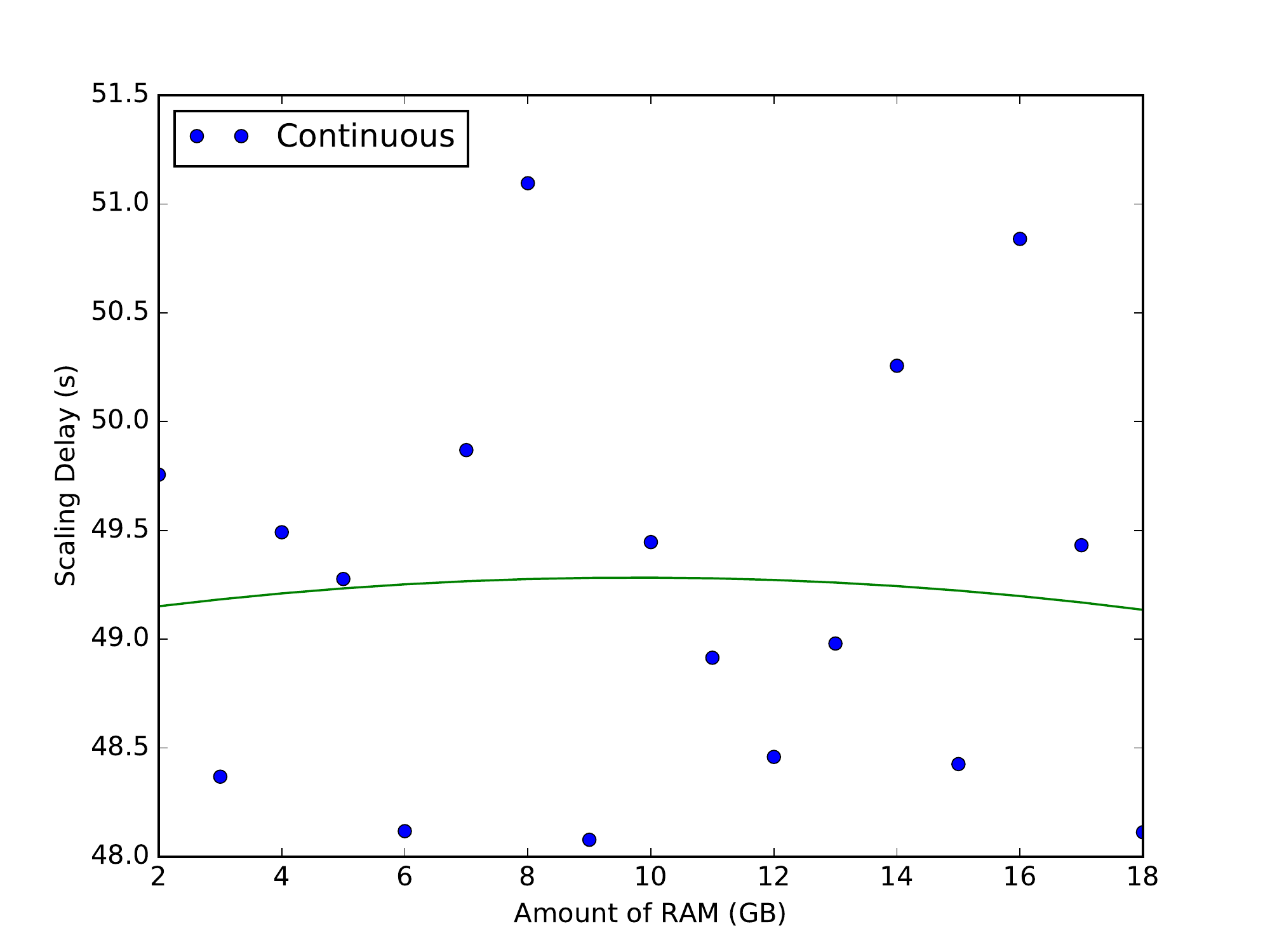}
	\caption{\label{fig:chap5:regressionmem_continuous} Second order polynomial function of mean RAM upscale time in continuous scenario}
\end{figure}
\begin{figure}
	\centering
	\includegraphics[scale=0.3]{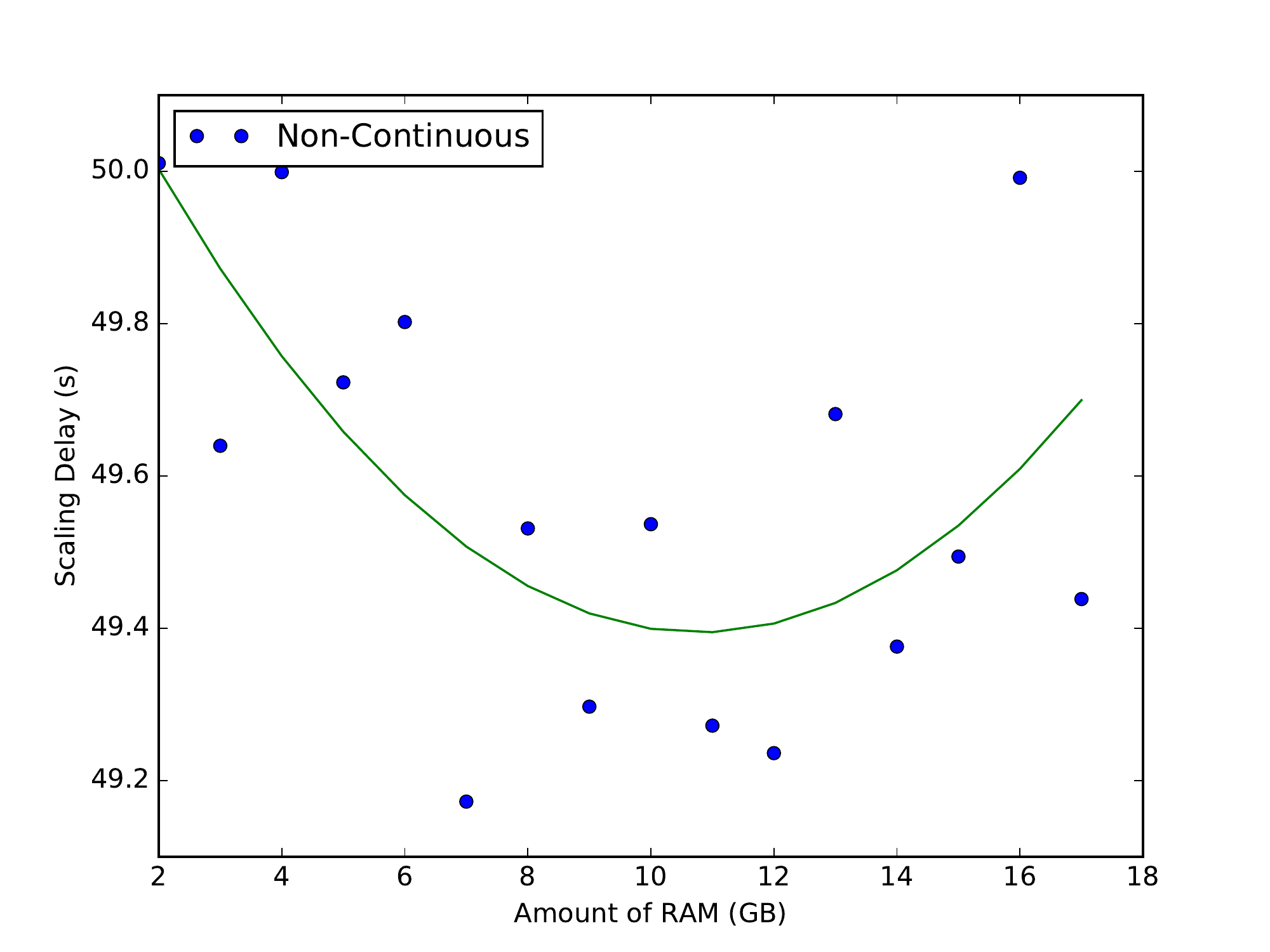}
	\caption{\label{fig:chap5:regressionmem_startfrom1}Second order polynomial function of mean RAM upscale time in non-continuous scenario} 
	
\end{figure}

\begin{figure}
	\centering
	\includegraphics[scale=0.3]{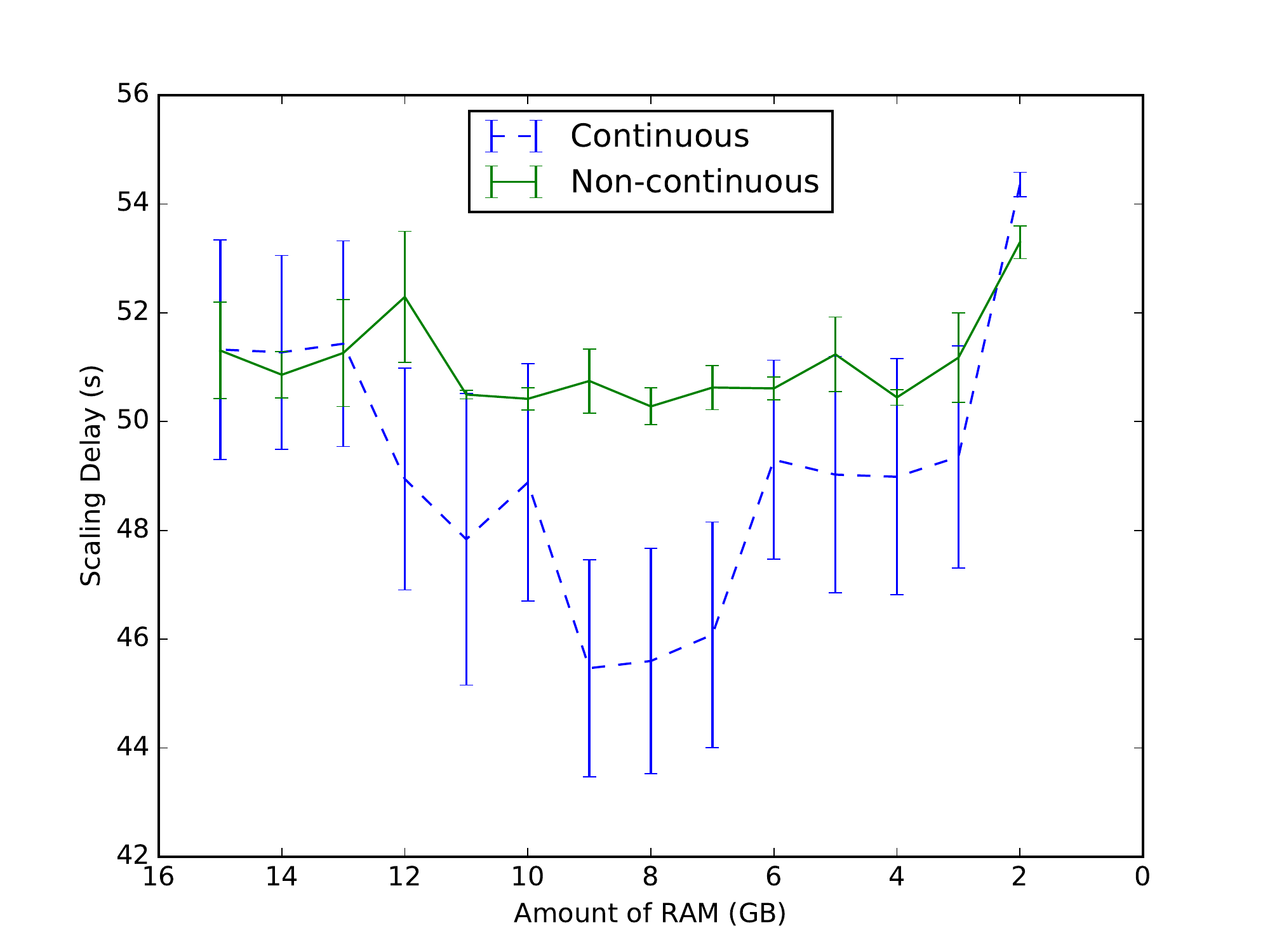}
	\caption{\label{fig:chap5:rev_mem_continuousVSstart} Mean RAM downscale delay as the size of the base VM increases. The standard error shows variations (error) in data}
\end{figure}
\begin{figure}
	\centering
	\includegraphics[scale=0.3]{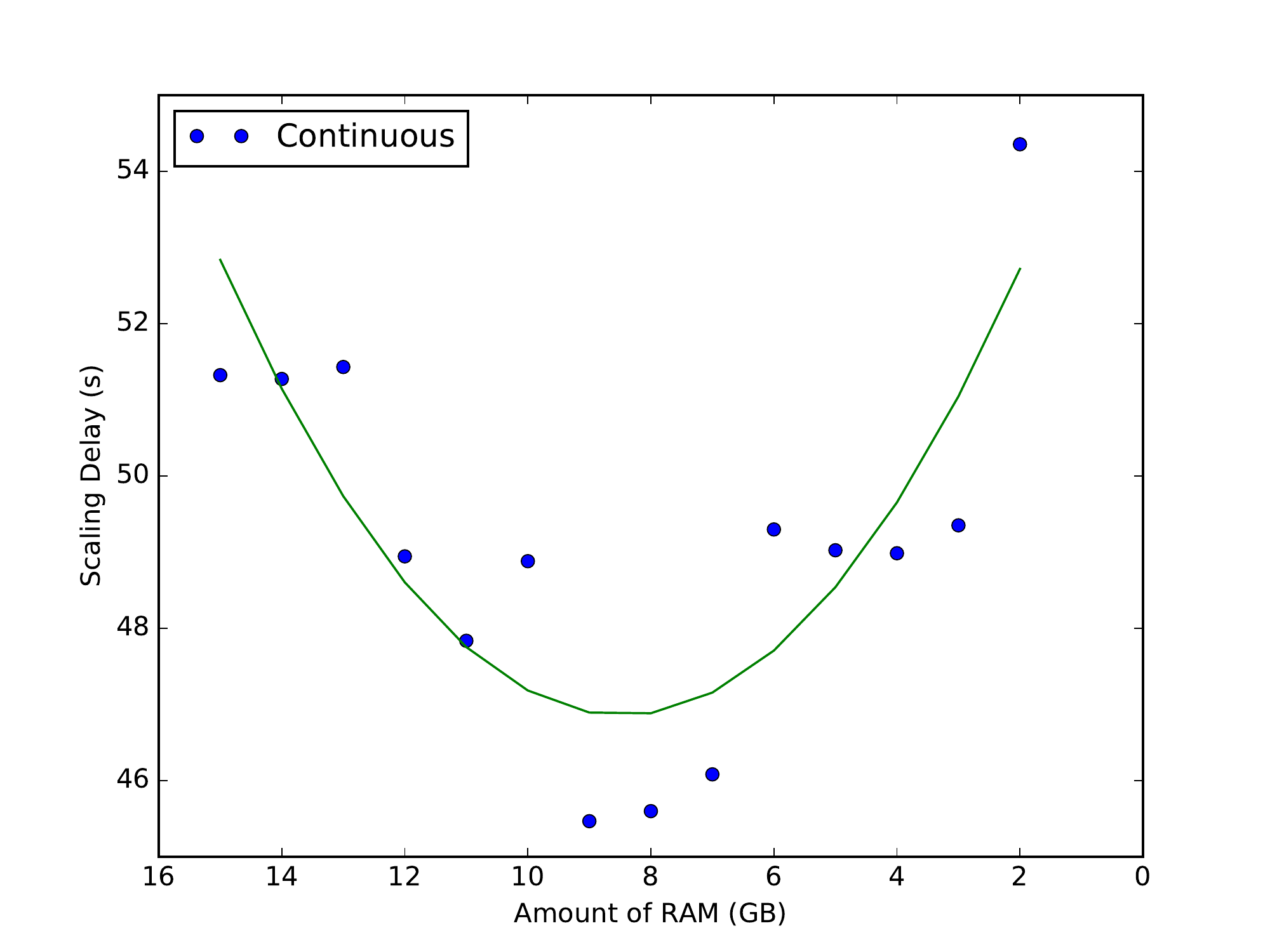}
	\caption{\label{fig:chap5:rev_regressionmem_continuous} Second order polynomial function of mean RAM downscale time in continuous scenario}
	
\end{figure}
\begin{figure}
	\centering
	\includegraphics[scale=0.3]{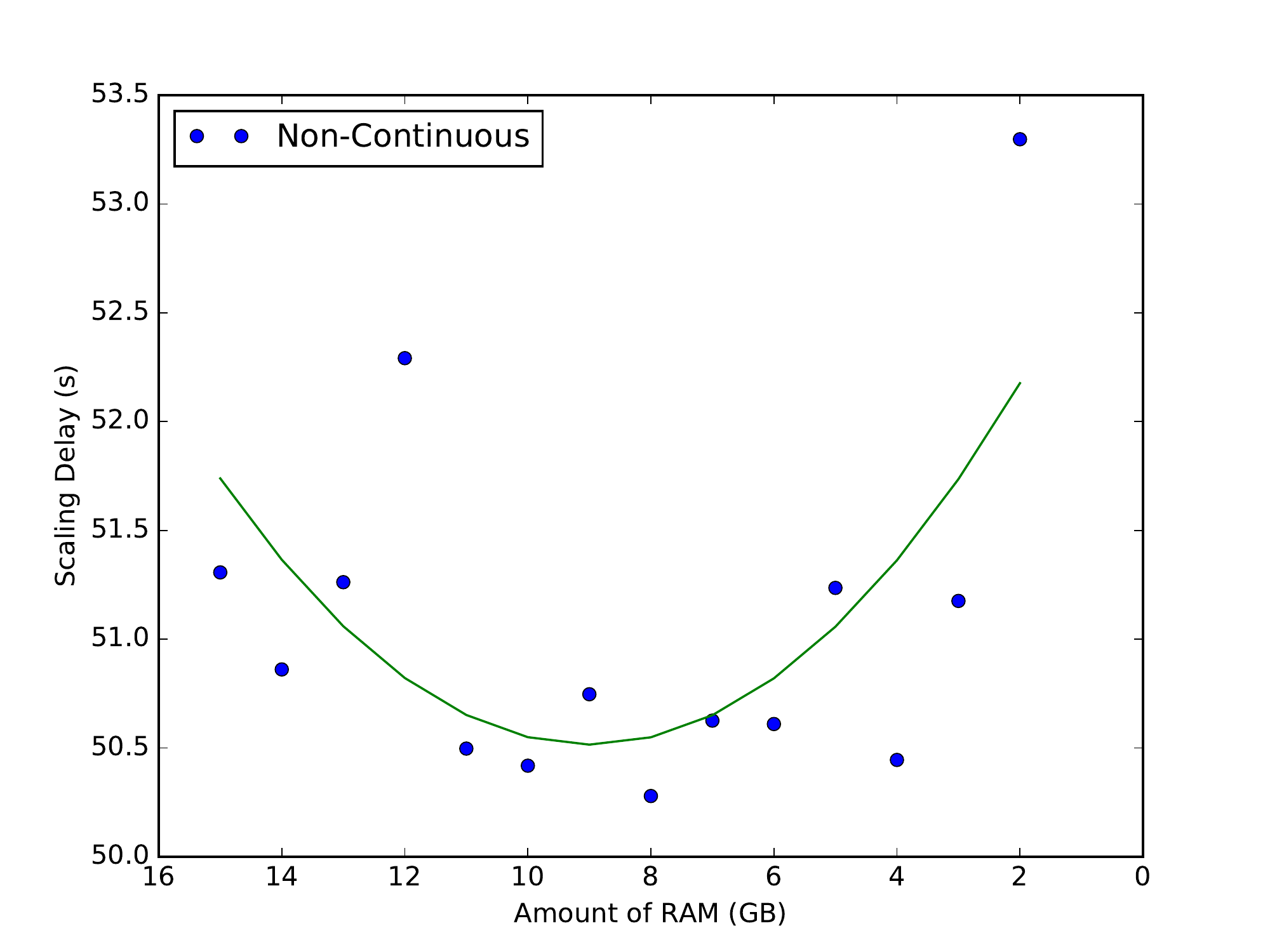}
	\caption{\label{fig:chap5:rev_regressionmem_startfrom1} Second order polynomial function of mean RAM downscale time in non-continuous scenario}
	
\end{figure}

Despite all of the above clear trends in results, when analysing resizing time delays when resizing RAM resource on VMs, no distinctive trends were found in both continuous and non-continuous scenarios, as shown in Figure \ref{fig:chap5:mem_continuousVSstart} for upscaling and \ref{fig:chap5:rev_mem_continuousVSstart} for RAM downscaling. However, this could be due to the fact that the test-bed's maximum memory resource limit is 18GB, and the sample amount is not large enough to see clear trends. The second order polynomial regression analyses show decreasing trends on both continuous and non-continuous scenarios, in Figure \ref{fig:chap5:regressionmem_continuous} and in \ref{fig:chap5:regressionmem_startfrom1}, in continuous and non-continuous upscaling scenarios respectively. Moreover Figure \ref{fig:chap5:rev_regressionmem_continuous} and Figure \ref{fig:chap5:rev_regressionmem_startfrom1} shows a second order polynomial regression analysis of both continuous and non-continuous downscaling of RAM resources.

From the lessons learned from above empirical VM resize performance analysis, it is evident that careful attention should be paid to the time delays when resizing CPU and storage resources of VMs, for real-time applications. Due to the significant delay incurred by the cloud framework (e.g. over 50 seconds in Figure \ref{fig:chap5:regressioncpu_continuous}), it may not be practical to auto-scale in real-time for delay-constrained applications. Instead, the controller may initiate the VM scaling process before the task execution (or by migrating tasks to another VM, during the resize process). The auto-scaling delay trends may vary depending on the underlying technologies (e.g. virtualization and physical resource management) that are deployed. This analysis brings awareness to the community that auto-scaling delays depend not only on the resource type but also on the auto-scaling method, i.e., upscaling, downscaling, continuous and non-continuous. We would like to stress that, when designing auto-scaling algorithms, CPU and storage resizing delay trends with various conditions of the VMs itself (e.g. VM size) and of the environment (e.g. total VM count in the cloud) that they are hosted in, should be considered. We have discovered that it is important to use both the knowledge of the VM itself and the cloud environment together at the same time, in future vertical scaling algorithms.

\section{Conclusion}
\label{sec:conclusion}

This paper has introduced a protocol that uses a simple unified packet header for both resource management and task offloading. A new logical controller that receives instantaneous monitoring information from both computing and communication sides and makes efficient resource management decisions on both C-RAN and mobile cloud for mobile task offloading. 

We conducted an analysis on scaling up and scaling down performances of MCC resources. The analysis shows that auto-scaling performances of all storage, CPU and RAM resources vary when scaling resources vertically. We can also conclude that auto-scaling in real-time is not practical due to high auto-scaling delay. The results also revealed that scaling time delay depends on the amount of resources that are added or removed from the VM at each step. One may conclude from the auto-scaling delay analysis that it is necessary to analyse auto-scaling performances of each cloud platform due to complex nature of today's cloud systems. Moreover, being aware of the scaling delay time trends may help to make effective auto-scaling decisions.

\section{Acknowledgment}
\label{sec:ack}

This work was supported by UK EPSRC NIRVANA project
(EP/L026031/1), EU Horizon 2020 iCIRRUS project (GA-
644526).



%
%

\bibliographystyle{ieeetr}
\bibliography{bib.bib}

\end{document}